\documentclass[%
 reprint,
superscriptaddress,
 amsmath,amssymb,
 aps,
twocolumn
]{revtex4-2}

\usepackage{graphicx}
\usepackage{dcolumn}
\usepackage{bm}
\usepackage{chemmacros}
\usepackage{float}
\usepackage{comment}
\usepackage{xcolor}
\usepackage[hidelinks]{hyperref}

\newcommand{\beq}{\begin{equation}}
\newcommand{\eeq}{\end{equation}}

\newcommand{\G}{\Gamma}
\renewcommand{\L}{\Lambda}

\renewcommand{\d}{\delta}

\renewcommand{\selectlanguage}[1]{} 

\begin{document}

\title{Evolutionary features in a minimal physical system:\\ diversity, selection, growth, inheritance, and adaptation}

\author{Guy Bunin}
\affiliation{Department of Physics, Technion-Israel Institute of Technology, Haifa 32000, Israel}
\author{Olivier Rivoire}
\affiliation{Gulliver, CNRS, ESPCI, Universit\'e PSL, 75005 Paris, France}

\begin{abstract}
We present a simple physical model that recapitulates several features of biological evolution, while being based only on thermally-driven attachment and detachment of elementary building blocks. Through its dynamics, this model samples a large and diverse array of non-equilibrium steady states, both within and between independent trajectories. These dynamics exhibit directionality with a quantity that increases in time, selection and preferential spatial expansion of particular states, as well as inheritance in the form of correlated compositions between successive states, and environment-dependent adaptation. The model challenges common conceptions regarding the requirements for life-like properties: it does not involve separate mechanisms for metabolism, replication and compartmentalization, stores and transmits digital information without template replication or assembly of large molecules, exhibits selection both without and with reproduction, and undergoes growth without autocatalysis. As the model is based on generic physical principles, it is amenable to various experimental implementations.
\end{abstract}

\maketitle

\section{\label{sec:intro}Introduction} 

Life is understood as both the product and the engine of evolutionary processes. In extant life forms, reproduction of individuals with heritable variation, which is required for Darwinian evolution, involves complex chemical and physical processes orchestrated by intricate biomolecules that are, moreover, themselves the product of evolution. The intricacy of these processes poses a significant challenge for conceptualizing life and understanding how it originated. This challenge motivates the development of simple physical models capable of mimicking biological evolution as an approach to elucidate the minimal requirements for life-like features.

The focus of many efforts has been the design of simple computational~\cite{von1966theory} or physical systems capable of reproduction. The latter range in scale from centimeters~\cite{PENROSE.1958rf} to microns~\cite{Zhou.2021} and nanometers~\cite{Kiedrowski.1986,Schulman.2012}. The designs typically require ingenious mechanisms and rely on external drives such as temperature cycling. Their common underlying principle is autocatalysis, where one element promotes the formation of other elements of the same type~\cite{Hanopolskyi.2021} or, more generally, where several elements collectively promote their reproduction~\cite{Hordijk.2017dm}. 

In any case, reproduction alone is not sufficient for Darwinian evolution: significant diversity must also be generated and inherited. Assemblies that reproduce by template replication, either in the form of heteropolymers~\cite{lincoln2009self}, supramolecular polymers~\cite{colomb2015exponential} or growing crystals~\cite{cairns1966origin,Schulman.2012}, are often taken as carriers of diversity. An alternative is to take as carriers of diversity compositional states, defined by the relative concentration of a set of chemical species~\cite{Segre.2000}. In this context, it has been proposed that evolution may operate through transitions between autocatalytic ``cores'', each supported by a different subset of molecules~\cite{Blokhuis.2020}. However, no experimentally feasible proposal has been made. Besides, it has been argued that additional processes, such as low-rate background reactions or the presence of compartments, may be necessary for evolution to occur by this process~\cite{Vasas.2012}. In light of these considerations, developing simple physical models capable of mimicking biological evolution remains a central and open challenge.

Here we propose a theoretical but physical and generic model that features directionality, diversity, selection, growth, inheritance and adaptation. It does not include autocatalysis, compartments, or assemblies into large molecules, demonstrating that they are not necessary for generating these evolutionary features.

\section{\label{sec:model}Model} 

We provide three levels of description for our model: a physical description that defines the fundamental components and their interactions, prescribing how these components bind and unbind due to thermal fluctuations; a detailed chemical description that lists the elementary reactions between the components; and a coarse-grained chemical description that retains only the key dynamical variables, from which the results presented in the figures are generated.

The three levels of description serve different purposes. The last, coarse-grained level is employed in numerical simulations to generate the results that we present. It is derived, under certain specified conditions, from the description of all elementary chemical reactions and their respective rates. This detailed chemical description, which can also be used to study thermodynamics, often serves as a starting point in models related to the origin of life. Here, we derive it from a more fundamental physical description. This approach ensures physical consistency, provides guidance for experimental implementation, and allows us to account for critical physical and geometric constraints that are not considered in chemical models based solely on thermodynamics -- constraints often associated with ``molecular and functional complexity''.
 
For example, these constraints explain why experiments with non-enzymatic autocatalysis often exhibit non-exponential growth~\cite{Kiedrowski.1986}, as one manifestation of these constraints is product inhibition~\cite{Sakref.2023tvn}. They also underlie the difficulty of evolving replicases large enough to allow faithful replication~\cite{mcginness2003search}, which led to the proposal of a threshold known as the error catastrophe~\cite{eigen1971selforganization}, again unexplained by thermodynamics. More generally, the purported need for large molecules to support evolutionary processes motivates many proposals that use molecular complexity as a biosignature~\cite{marshall2021identifying}. By defining our model at the physical level, we make explicit what we mean by a minimal and physical model, namely one based only on rigid objects that does not require or assume any assembly or any enzyme-like mechanism.

\subsection{Physics and Chemistry}

Inspired by a model in theoretical ecology showing how networks of inhibitory interactions lead to many stable states, and directional dynamics over long time scales with regimes of spatial expansion~\cite{Bunin.2021}, we define a chemical system with a similar network of inhibitory interactions between constituents. 
The stable states of the chemical system are defined by the concentrations of $N$ reactants $A_i$, $i=1..N$. The production of each $A_i$ is promoted by a catalyst $C_i$, which can be inhibited by the presence of certain other $A_j$. This can lead to multistability, where certain reactants are ``active'', i.e. found at high levels, and prevent others from being catalyzed at significant rates. This multistability generally requires a nonlinear dependence of the inhibition of $C_i$ on the concentration of the inhibiting $A_j$~\cite{Ferrell.2001}. One mechanism often encountered in biological systems is cooperativity, where inhibition involves two or more $A_i$ that interact non-additively. Another mechanism is inhibitor ultrasensitivity~\cite{Ferrell.1996}, which requires only the presence of another molecule $B_i$ that strongly binds to $A_i$. We adopt this simpler mechanism where the nonlinearity arises from the necessity for $A_i$ to first saturate $B_i$ before effectively inhibiting $C_j$.

\begin{figure}[t]
\centering
\includegraphics[width=\linewidth]{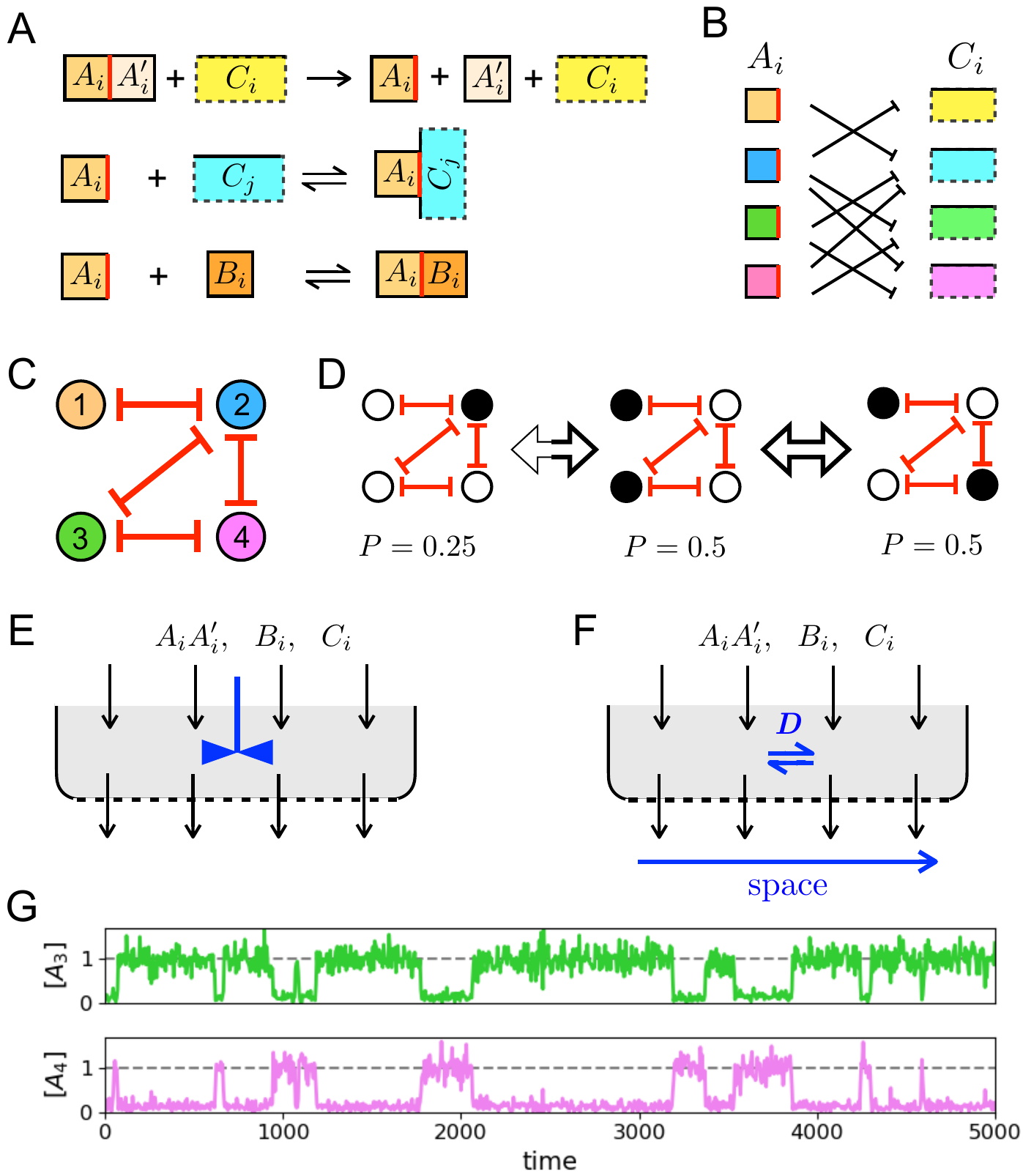}
\caption{{\bf A.} We consider $N$ reactions $(i=1..N)$ involving the dissociation of dimers $A_iA_i'$ into monomers $A_i$ and $A_i'$, catalyzed by $C_i$. The product $A_i$ can bind to and inhibit the catalyst $C_j$ of another reaction $j$. In addition, $A_i$ can bind to $B_i$ to form a dimer $A_iB_i$. {\bf B.} Inhibition is reciprocal: if $A_i$ inhibits $C_j$, then $A_j$ inhibits $C_i$. {\bf C.} The network of reciprocal inhibition forms a graph where nodes represent reactions $i$, and edges reciprocal inhibition. {\bf D.}~Representation of the 3 stable states associated with the graph in C. By default, a node is active unless repressed by an active neighboring node. In a stable state, a node is active (in black) if and only if all the nodes to which it is connected are inactive (in white). The productivity $P$ of a state is the fraction of active nodes. Transitions between states exhibit directionality, predominantly occurring from low to high productivity states. {\bf E.} The reactions take place in an open reactor where $A_iA'_i$, $B_i$ and $C_i$ are continuously injected and where all species are continuously diluted out. When the system is well mixed, the reactants are homogeneously distributed throughout the reactor. {\bf F.}~Injection and dilution are homogeneous in space, but heterogeneities can arise from stochastic fluctuations and the limited diffusion of all species along a one-dimensional direction. {\bf G.} Example trajectory for the concentrations $[A_i]$ associated with nodes $i=3$ and 4 of the graph in C in the well-mixed case, showing stochastic switches between the two states with highest productivity $P=0.5$ ($[A_1]\simeq 1$ and $[A_2]\simeq 0$ remain roughly constant, see Fig.~S1).
\label{fig:scheme}}
\end{figure}

The reaction that produces $A_i$ could be of various types. To emphasize that neither large molecules nor assembly are important, we consider $A_i$ to result from the dissociation of a dimer $A_iA'_i$ (Fig.~\ref{fig:scheme}A). $B_i$ is defined as another monomer that forms a dimer $A_iB_i$ when it interacts with $A_i$. This interaction is through the interface by which $A_i$ interacts with both $A_i'$ and $C_j$, so that $B_i$ and $C_j$ interact with $A_i$ but not with $A_iA'_i$ (Fig.~\ref{fig:scheme}A and Appendix~A). The catalyst $C_i$ accelerates the cleavage reaction $A_iA'_i\to A_i+A'_i$ without being consumed. In biology, most catalysis is mediated by enzymes, which are large molecules. To implement an elementary form of non-enzymatic catalysis, we define $C_i$ as a rigid body with two interacting sites for binding to $A_i$ and $A'_i$, which achieves catalysis provided that the distance between these two interacting sites and their affinity are in appropriate ranges~\cite{Rivoire.2020,Munoz-Basagoiti.2023}. 

An essential aspect of the model is the topology of the network of inhibitory interactions. Each $A_i$ does not inhibit every $C_j$ but only a small number (small compared to $N$), and the inhibition is reciprocal: if $A_i$ inhibits $C_j$, then $A_j$ also inhibits $C_i$ (Fig.~\ref{fig:scheme}B). This choice is motivated by general results in statistical mechanics and theoretical ecology showing that such sparse reciprocal interactions allow one to obtain a large number of stable states~\cite{barbier2013hard,Fried.2017,Bunin.2021}. The structure of the reciprocal inhibitory interactions is summarized in an undirected graph, where each node represents a reaction $i$ (Fig.~\ref{fig:scheme}C). Below, we take a random-regular graph with connectivity $c=3$, that is, randomly chosen from graphs where each node is connected to exactly three other nodes.

\subsection{Effective reaction kinetics}

We study the reactions in the setting of an open system where the substrate $A_iA'_i$ for the production of $A_i$, the binders $B_i$, and the catalysts $C_i$ are supplied externally at a constant flux, and where all chemical species are continuously diluted out at a constant rate. If the system is well mixed, this corresponds to a chemostat setup (Fig.~\ref{fig:scheme}E). However, to obtain reproduction in our model, space is essential, and we therefore consider that each chemical species can diffuse in space with a diffusion constant that we take to be the same for all species  (Fig.~\ref{fig:scheme}F). The supplied chemicals are delivered uniformly throughout space. For simplicity, we assume a one-dimensional space with periodic boundary conditions. Finally, the discreteness of the molecules induces stochasticity, also known as demographic noise in the context of evolutionary dynamics.

Such a system is described mathematically by a set of coupled reaction-diffusion Langevin equations (Appendix~D). To simplify the presentation and analysis, we consider a limit where time-separation leads the dynamics to be effectively described by a set of only $N$ equations. In this limit, the inhibition of $C_j$ by $A_i$ occurs on a faster timescale than the catalysis of $A_i$ by $C_i$, and the binding of $B_i$ to $A_i$ on an even faster timescale (Appendix~B). A large well-mixed system is then simply described by
\beq\label{eq:dyn}
\partial_{t}[A_i]=\frac{\L_i}{1+\Gamma\sum_{j\sim i}\max(0,[A_j]-[B_j])}-\d [A_i] 
\eeq
where $[A_i]$ and $[B_i]$ represent the concentration of the elements $A_i$ and $B_i$, including those in the form of a dimer $A_iB_i$, but excluding the dimers $A_iA_i'$. In this limit, $[B_i]$ is constant and can be treated as a model parameter. The first term of the equation represents inhibited catalysis, where the sum in the denominator is over all the $A_j$ that can potentially inhibit $C_i$, and the second term represents dilution at a rate $\d$. We first consider $\L_i$ to be the same for all $i$, which corresponds to an uniform supply of each substrates $A_iA_i'$ and an equal catalytic efficiency of each catalyst $C_i$. Besides the graph of inhibitory interaction defining the relationships $i\sim j$, the equation has the following independent parameters: the inhibitory strength $\G$, the concentration $[B_i]$ which we take to be independent of $i$, corresponding to an uniform supply of all $B_i$, and the $\Lambda_i$s (the dilution rate $\d$ can always be absorbed into a redefinition of the time unit). This equation is generalized to include diffusion and demographic noise to describe the stochastic dynamics for the local concentration $[A_i](x,t)$ of molecules $A_i$ at each spatial location $x$ and time $t$ (Appendix~D). Diffusion is quantified by a diffusion constant $D$, and demographic noise by a parameter $\omega$ that scales with the inverse of the square root of the volume (Appendix~C). We simulate these equations with a standard Euler-Maruyama algorithm, adding only a constraint to force $[A_i](x,t)$ to be non-negative (Appendix~F).

Unless otherwise stated, we fix the following parameters: $N=50$, $\L_i=1$, $\G=10$, $\d=1$, $[B_i]=0.25$, $\omega=0.1$. We discretize space in $M=10$ or $M=100$ cells of unit length, while keeping a constant diffusion constant $D=1$. To simulate a corresponding well-mixed system, we divide $\omega$ by $\sqrt{M}$ and then set $M=1$. The dynamics being stochastic whenever $\omega>0$, different realization of the dynamics are obtained when starting from the same initial condition, which by default we take to be  $[A_i](x,t=0)=0$ for all $i$ and $x$.

\section{\label{sec:results}Results}

Simulations of the spatio-temporal dynamics described above show that variables spend most of the time fluctuating around two values, with abrupt switches between them: $A_i$ is either ``active'', $[A_i]\simeq \L_i/\delta$, or ``inactive'', $[A_i]\simeq 0$ (Fig.~\ref{fig:scheme}G). These dynamics can be understood as fluctuations around stable states of the deterministic dynamics without demographic noise. Stable states are configurations of active and inactive $A_i$, where $A_i$ is active if and only if all the $A_j$ to which it is connected by inhibitory interactions are inactive (Fig.~\ref{fig:scheme}D). In graph theory, these states are known as the maximal independent sets of the inhibition graph. Enumerating them in a well-known combinatorial problem~\cite{papadimitriou1998combinatorial}. Noise-driven jumps between the basins of different stable states involve switches between active and inactive values for a number of variables. The distribution of concentrations is therefore essentially bimodal (Fig.~S4A), although the concentrations may take intermediate values at the spatial boundaries separating two stable states.

\begin{figure}[t]
\centering
\includegraphics[width=\linewidth]{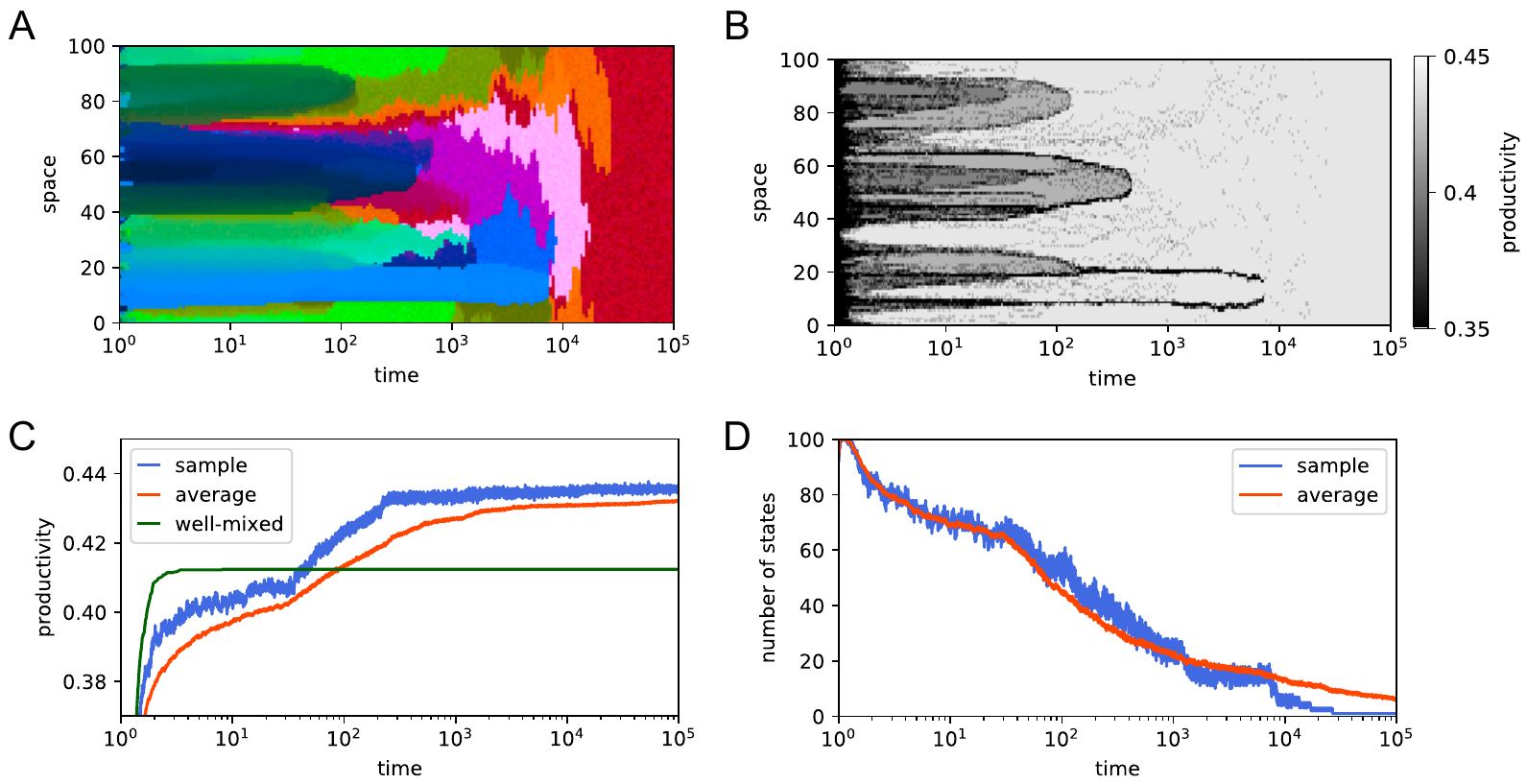}
\caption{{\bf A.} A representation of one trajectory in space and time using a low-dimensional projection of the relative concentrations $[A_i]$ to visualize different compositions by different colors (Appendix~F). Here the inhibition graph is a random regular graph with $N=50$ nodes and connectivity $c=3$. Space is divided in $M=100$ cells and time is represented in log-scale (see Fig.~S2 for other sample trajectories from the same inhibitory graph and Fig.~S3 for sample trajectories from other graphs). {\bf B.} For the same trajectory, productivity, i.e, fraction of active $A_i$. {\bf C.} Mean productivity across space. For comparison, average over 20 different trajectories obtained with exactly the same parameters, and the result for a corresponding well-mixed system. The sample trajectory reaches the maximum productivity ($P=22/50=0.44$) for the chosen inhibitory graph. {\bf D.} Evolution of the number of different stable states. The average value shows that while the sample trajectory ends up in a single state, this is not always the case (see Fig.~S2).\label{fig:dir}}
\end{figure}

An illustration of one realization of the process is shown in Fig.~\ref{fig:dir}A. In this representation, borrowed from analytical chemistry~\cite{gardner2020self}, different colors represent different chemical compositions, with similar colors indicating similar compositions (Appendix~F). At each space and time point, we identify the nearest stable state (Appendix~F), which is a steady state for the deterministic dynamics without demographic noise. This nearest stable state essentially amounts to considering $A_i$ as active or inactive depending on whether it is closer to $0$ or to $\L_i/\d$ (Fig.~S4B). 

\subsection{Extended directional evolution and reproduction}

We reproduce an observation first made in the context of an analogous Lokta-Volterra system with sparse inhibitory interactions~\cite{Bunin.2021}: changes occur on timescales that far exceed any microscopic timescale ($T=10^5$ in Fig.~\ref{fig:dir} while all parameters in Eq.~\eqref{eq:dyn} are of order 1; see also Figs.~S2-S3 for other sample trajectories with the same or different inhibitory graphs). Unlike the Lokta-Volterra system, however, our model does not have a Lyapunov function. Nevertheless, these changes are directional: the system evolves towards states of increasing ``productivity'', where productivity of a state is the fraction of active variables $A_i$ in this state (Fig.~\ref{fig:dir}B-C). Configurations sampled through the dynamics are not strictly stable states but the productivity of a configuration can be defined by considering the nearest stable state to that configuration. A directional evolution toward higher productivity is also obtained in well-mixed systems (Fig.~\ref{fig:dir}C and Fig.~S5), in which case it can be understood formally. For well-mixed systems with strong inhibition (large $\G$) and low noise (small $\omega$), transitions between states are indeed entirely driven by fluctuations that reduce the concentrations of active variables, which then allows previously inactive neighboring variables to become active. The transition from a high-productivity state to a low-productivity state requires more of these fluctuations and is therefore less probable than the reverse transition (Appendix~E). This asymmetry in transition probabilities results in directional dynamics that favor progression toward higher productivity states over time.

When mixing is not instantaneous, allowing for variation in space, the dynamics is more complex and this increase in productivity is accompanied  by a spatial expansion of some states at the expense of others, which is a form of reproduction. 
This is evidenced by a decrease in the number of different states as a function of time (Fig.~\ref{fig:dir}D).

\subsection{Diversity and selection}

\begin{figure}[t]
\centering
\includegraphics[width=.9\linewidth]{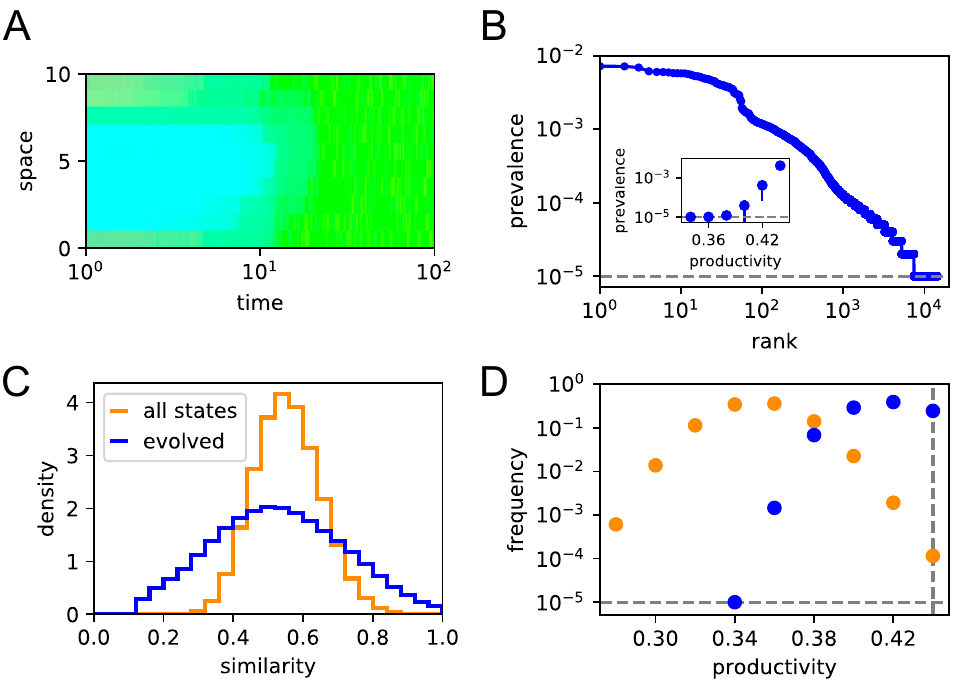}
\caption{{\bf A.} Illustration of the evolution of a system of size $M=10$ over a time $T=10^2$, starting from all $[A_i]=0$.
{\bf B.}~When this evolution is repeated $10^5$ times, some final states are reached more often than others: the most common are found in nearly 1\% of the samples (prevalence $\sim 10^{-2}$) while the least common are found only once (prevalence $10^{-5}$). Inset: Mean prevalence of states with given productivity, with error-bars indicating the standard deviation of the distribution of prevalences. {\bf C.} Distribution of pairwise similarities between states, measured by the fraction of active or inactive $A_i$ that two states share (in blue). For comparison, distribution of similarities between all stable states (in orange). {\bf D.} Distribution of productivity in evolved states (in blue) versus all stable states (in orange). The horizontal dashed line at frequency $10^{-5}$ marks the sampling limit given the $10^5$ sample trajectories while the vertical dashed line marks the maximum productivity for this system.\label{fig:div}}
\end{figure}

When starting multiple replicate trajectories from the same all-zero initial condition (Fig.~\ref{fig:div}A), a large number of different stable states are attained in a finite time (Fig.~\ref{fig:div}B). Measured by the Shannon diversity, the number of states is in the thousands (in this instance, around 3000 states; see Appendix~F). Moreover, these states are very different from each other (Fig.~\ref{fig:div}C).
However, this diversity is far less than the total number of stable states, which for the particular inhibitory graph used in Figs.~\ref{fig:dir}-\ref{fig:div}, is $\mathcal{N}\simeq 4.8\ 10^5$. More generally, for random-regular graphs of connectivity $c=3$, the number of stable states scales exponentially with the number $N$ of nodes, as $\mathcal{N}\sim e^{Ns}$ with $s\simeq 0.26$ (Fig.~S7), and similarly for other random sparsely-connected graph ensembles~\cite{Fried.2017}.

As further evidence that selection is taking place, the subset of states that are reached most often are those with the highest productivity (Fig.~\ref{fig:div}D), which is consistent with the overall growth of productivity (Fig.~\ref{fig:dir}C). Productivity is thus analogous to fitness: current states with the highest productivity contribute more to future states. 
This is seen in the dynamics at intermediate and late times, $t\gtrsim10$, where states of higher productivity are more likely to expand in space, leading to an increase in the average fitness (Fig.~\ref{fig:dir}A). This analogy with fitness is further supported by the analysis of competitions below.
Note, however, that productivity increases even in well-mixed systems in which reproduction by spatial expansion cannot occur (Fig.~S5), which represents a form of selection even in the absence of reproduction.

\subsection{Heredity and competition}

In well-mixed conditions, successive states become increasingly similar over time, with the rate of change between states gradually decreasing. This trend can be interpreted as an evolution towards more accurate inheritance (Fig.~S5). Similar features are observed in systems with limited diffusion, as demonstrated by temporal correlation functions (Fig.~S6). Additionally, in systems with limited diffusion, certain states expand at the expense of others, exhibiting a form of reproduction that also involves inheritance. This spatial inheritance, which is \emph{a priori} distinct from the increasing correlation in time, is quantified through spatial correlation functions (Fig.~S6).

\begin{figure}[t]
\centering
\includegraphics[width=.9\linewidth]{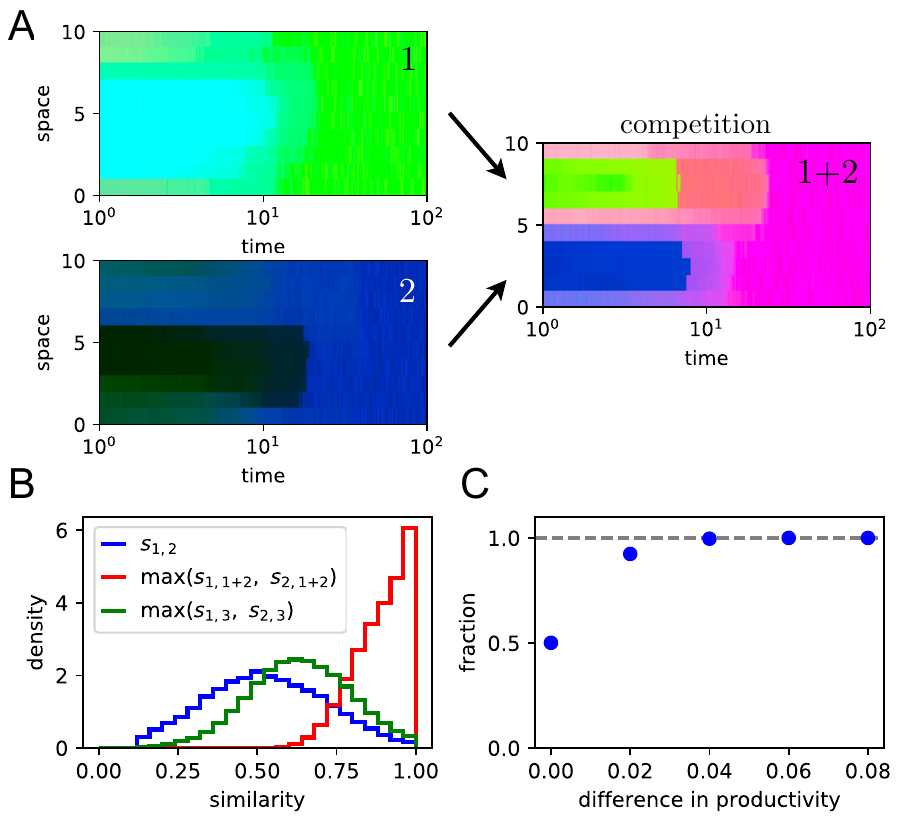}
\caption{{\bf A.} Two compositions obtained after evolution from the all-zero state (Fig.~\ref{fig:div}A) are competed against each other by juxtaposing them in space. {\bf B.} Distribution of pairwise similarities between the state resulting from the competition ($1{+}2$) and the closest of the two competing states ($1$ or $2$). For comparison, we also show the closest similarity of another independently evolved state $3$ to either $1$ or $2$, and the similarity between 1 and 2. {\bf C.} 
Fraction of times that the final state is more similar to the state with the higher productivity of the two initial competing states, as a function of the productivity difference between the two initial competing states.
\label{fig:compet}}
\end{figure}

To push this idea further, we consider competitions between two evolved states that are obtained by the dynamics of the previous subsection. The competition consists of initializing a system with these two states, each occupying half of the one-dimensional state space, and recording the state reached after a finite time (Fig.~\ref{fig:compet}A). This usually results in a state that is different from either of the two competing states, but still much closer to one of them than to any other evolved state (Fig.~\ref{fig:compet}B). When the two competing states have different productivity, this closest state is almost always the competing state with greater productivity (Fig.~\ref{fig:compet}C), a result consistent with viewing productivity as a form of fitness.

The concepts of heredity and productivity are distinct: heredity refers to persistent temporal correlations, while productivity refers to the number of active nodes. In our model, however, they increase concomitantly. In the well-mixed setting, at the low-noise limit (Appendix~E), our model maps to a mean-field glass model~\cite{barbier2013hard}, and this coupling reflects the well-known phenomenon of aging, where temporal correlations increase while energy decreases~\cite{arceri2022glasses}. Under limited diffusion, these effects are additionally coupled to spatial expansion (growth) and behavior under competition, features not typically associated with glassy systems.

\subsection{Environments and adaptation}

In evolutionary biology, adaptation is contingent on the particular environment in which organisms evolve. Here, the environment comprises the substrates $A_iA'_i$ that are injected. We have so far assumed that all substrates are injected at the same rate, but different environments can be defined by assuming that they are injected at different rates, which corresponds to taking $\L_i$ to depend on $i$ in Eq.~\eqref{eq:dyn}.

\begin{figure}[t]
\centering
\includegraphics[width=.5\linewidth]{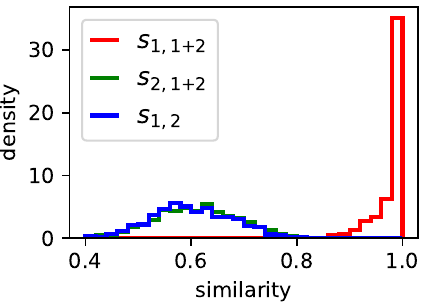}
\caption{An environment is defined by the influx rates of the substrates $A_iA'_i$ from which the $A_i$ originate (Fig.~\ref{fig:scheme}A). Formally, it is specified by a $N$-dimensional vector with components $\L_i$ (see Eq.~\eqref{eq:dyn}). Taking these components uniformly at random in $[0,1]$ to define two environments $\vec{\L}^{(1)}$ and $\vec{\L}^{(2)}$, we evolve two systems, one in each environment, and let them compete in the environment $\vec{\L}^{(1)}$. The state resulting from this competition is very similar to the state of the system that evolved in the same environment ($s_{1,1+2}$ close to 1), and very dissimilar to the other state that evolved in a different environment ($s_{2,1+2}$ comparable to $s_{1,2}$). This is consistent with an adaptation of that system to its environment ($M=10$, $T=10^2$ as in Fig.~\ref{fig:compet}). \label{fig:adapt}}
\end{figure}

If we take two states resulting from an evolution in two different environments and compete them in one of the two environments, we observe that the resulting state is most similar to the state that evolved in the same environment, consistent with the notion that this state is adapted to its environment (Fig.~\ref{fig:adapt}). We also note that in a well-mixed system, the expression for the productivity generalizes to provide an environment-dependent notion of fitness (Appendix~E).

\section{Discussion and conclusion}

While nearly every specific feature of living systems has a counterpart in one or more non-living systems, understanding what is required for a physical system to exhibit the full range of features observed in living systems remains an open question. Here we focus on a subset of evolutionary features: directionality, diversity, selection, growth, inheritance and adaptation. In Darwinian evolution, diversity, differential reproduction and inheritance form the foundations from which adaptation emerges through natural selection~\cite{darwin1859origin}. Our model follows a different logic. Selection is present even in the well-mixed case, without spatial growth (Fig.~S5).  Spatial variation, when mixing is not instantaneous, introduces a new resource--that is, space--over which competition can occur. This spatial resource transforms selection into a more Darwinian process in which more productive states grow at the expense of less productive ones (Fig.~\ref{fig:dir}). This only begins after some time, consistent with the view that other forms of evolutionary adaptation, which do not require reproduction for selection and adaptation, may precede Darwinian evolution~\cite{Nowak.2008,smith2016origin,Wong.2023}. Our model also does not consider metabolism, replication and compartmentalization as separate processes that would need to be integrated~\cite{ganti_chemoton}. Instead, the same network of inhibitory reactions keeps the system in a state of disequilibrium, supports hereditary information and induces spatial boundaries. 

The essence of our model is an open dynamical system with a large number of stable steady states between which transitions occur stochastically, and which may expand in space. The analogy comparing states to biological species and state transitions to mutations in this context is not new and has been articulated particularly by Decker, who called such systems bioids~\cite{Decker.1974}. More generally, our model belongs to the broader class of models where the units of selection are relative molecular concentrations rather than specific (autocatalytic) molecules. Another model in this class is the GARD (Graded Autocatalysis Replication Domain) model, which describes autocatalytic growth of amphiphile assemblies~\cite{Segre.2000}.

Our model differs from these and previous models with compositional states in several ways. (1)  It is based on simple rigid elements that attach and detach and does not require the presence or assembly of any large molecule. (2) It exhibits temporal directionality, with an emerging notion of ``fitness'' that predicts the outcome of competitions between states. (3) It exhibits ``unlimited inheritance'', i.e., the capacity to be in a number of states much larger than the number of states that it samples at any given time~\cite{maynard_smith_major_1995}. (4) It does not involve autocatalysis.

The notion of autocatalysis is central to almost all previous models displaying evolutionary features, including those not based on compositional states. Autocatalysis is even often taken as synonymous with reproduction for a chemical system~\cite{calvin1969chemical}, with a distinction made between direct autocatalysis, where a molecule catalyzes its own formation, and reflexive (auto)catalysis, where a network of reactions is involved~\cite{Calvin.1956}. Several proposals have been made to formalize autocatalysis and allow a non-tautological association with growth and reproduction. Most of these formalizations focus primarily on necessary stoichiometric conditions~\cite{Hordijk.2017dm,Blokhuis.2020}, i.e., conditions that a network of reactions must satisfy independently of kinetic rates or initial conditions. Our model does not satisfy these stoichiometric conditions. The feedback mechanism leading to multistability is based on  inhibition rather than autocatalysis, and reproduction is observed only when spatial diffusion is considered: a well-mixed system that is defined by the same reactions shows mutations but neither growth nor reproduction. Unlike models based on stoichiometric autocatalysis, different states of our system are underlined by the same ``core'' of reactions~\cite{Vasas.2012,Blokhuis.2020}, and stochastic transitions between states stem from fluctuations in the small numbers of molecules without requiring any additional ``background'' reactions. The mechanism that generates the different states is more akin to symmetry breaking, where a common set of constraints admits several different but related solutions~\cite{anderson2018broken}.

Many chemical and physical models intended to imitate evolutionary dynamics are designed to alternate between phases of growth and fission, the latter typically triggered by external mechanical or thermal drives~\cite{Segre.2000,Schulman.2008,He.2017,Adamski.2020,Zhou.2021}. 
This aims to achieve an overall exponential growth by enabling each separated product to regrow independently. While we could introduce advection to achieve a similar effect, we prefer to emphasize that exponential growth is not necessary to obtain the evolutionary properties that we describe, particularly exclusion by selection. Furthermore, a variant of our model with an appropriate spatial connectivity pattern through which diffusion occurs can exhibit exponential growth without alternation (Fig.~S8). Regarding exclusion, while exponential growth provides one mechanism through the competitive principle~\cite{Szathmáry.1989}, it is not the only way, as illustrated in Fig.~\ref{fig:dir}A and demonstrated in other contexts~\cite{Sakref.2023tvn}. Selection between states in our model also does not arise from competition for the supplied substrates $A_iA'_i$, which are never exhausted. Instead, selection arises from the relative stability of states and, when diffusion is limited, from an additional competition for space.

Our model relies on two fundamental properties  of its building blocks: catalysis and specificity. We derive catalysis from fundamental physical principles, thus ensuring that it is a simple form of catalysis that can be implemented by rigid (inert) substances. Despite the inherent limitations of such simple catalysts, we demonstrate that any parameter values of the coarse-grained model are, in principle, achievable. Showing that the model can be realized without enzyme-like catalysts obviates the need to account for the emergence of such complex molecules. This contrasts, for example, with models with template replication of long heteropolymers mediated by polymerase-like catalysts, where the involvement of complex molecules to ensure accurate replication is thought to impose a constraint in the form of an error threshold~\cite{eigen1971selforganization}. This also contrasts with models that require temperature cycles or other forms of external drives. Note that we could also make the catalyst $C_i$ non-flowing, like minerals in some origin of life scenarios, and our results would not change. Specificity of the interactions is critical and requires a minimal form of complexity. Experimentally, this could be implemented with heteropolymers such as RNAs, which can form many specific interactions even with short sequences~\cite{santalucia1998unified,semizarov2003specificity}. Other assumptions we have made, such as that all types are strictly equivalent, are for convenience and are not necessary.

Genericity is a key feature of our model. The basic dynamical mechanism requires a set of catalyzed reactions, in which the product of some reactions inhibits other reactions in a sparse and reciprocal manner. We have presented an implementation of this mechanism with the dissociation of dimers into monomers, but the reverse, binding reaction could be considered to achieve the same phenomenology. Autocatalytic reactions could also be considered, where there is no distinction between the catalyst and the product of the reactions, thus reducing the number of different elements. Beyond chemistry, the same principle drives multistability in models of ecological dynamics~\cite{Bunin.2021} and gene regulation~\cite{gardner2000construction}. Indeed,  the same mechanism of reciprocal inhibition is obtained in models where $A_i$ represent individuals of a species born of other individuals of the same species, and inhibition takes the form of inter-species competition, or where they represent transcription factors transcribed by genes $C_i$, and inhibition takes the form of gene repression.
In this latter context, a mechanism for biased jumps between two states has been proposed and demonstrated, where noise in gene expression allows for adaptive growth in the absence of explicit regulator~\cite{furusawa2008generic}. There, however, the bias stems from cellular growth and division, with no analogy to the mechanism of chemical growth by diffusion that our model exhibits.
Above, we considered reciprocal inhibitory interactions, but the qualitative features are maintained even with some degree of asymmetry~\cite{Bunin.2021} (where the product of one reaction inhibits another reaction but not vice-verse). We also assumed stochasticity to arise from the small numbers of molecules but similar results could be obtained with large numbers of molecules if an another source of stochasticity drives the transitions between states.
Owing to the simple physical and generic mechanism on which it is based, our model can potentially be implemented in a variety of substrates at the molecular or colloidal level.

The model certainly lacks many characteristics associated with life, especially open-endedness (the potential for unlimited growth in complexity) and self-reference (no clear distinction between states and dynamical rules)~\cite{goldenfeld2011life}: the state space is large but predefined and finite, and optimal productivity can be reached after a long but finite time. Evolution does not stop there, since further transitions occur between states, but not further adaptation. However, one can imagine generalizations in which, for example, a state in which both $A_i$ and $A_j$ are present in high concentration leads to the formation of substantial amount of a new dimer $A_iA_j$ which can catalyze the formation of another $A_k$ as $C_k$ does, leading to a new state different from all those we have described, thus expanding the state space. This and other generalizations of the model are interesting directions for future work.

\acknowledgments

We are grateful to Philippe Nghe and Yann Sakref for their comments on an earlier draft.

\end{document}


\title{Evolutionary features in a minimal physical system:\\ diversity, selection, growth, inheritance, and adaptation}

\author{Guy Bunin}
\affiliation{Department of Physics, Technion-Israel Institute of Technology, Haifa 32000, Israel}
\author{Olivier Rivoire}
\affiliation{Gulliver, CNRS, ESPCI, Universit\'e PSL, 75005 Paris, France}

\maketitle

\appendix

\setcounter{page}{1}

\onecolumngrid

\centerline{\bf \Large Supplementary Information}

\section{Microscopic model}\label{app:chem}

We derive the chemical reactions shown in Fig.~1 from physical principles, starting from a description of the geometry and interaction potentials of fundamental elements. This ensures full consistency not only with thermodynamics but also with the constraints of elementary catalysts. The interaction potentials are simplified to two parameters, a forward and a backward activation energy, which control the kinetic rates through the Arrhenius law. This approach follows previous work showing that it accounts for properties obtained from molecular dynamics simulations with spherical particles at sufficiently low temperatures~\cite{Munoz-Basagoiti.2023}. For illustrative purposes, we draw the basic elements as squares or rectangles in two dimensions, with interaction patches at specific locations on their surface controlling the strength and specificity of their interaction with other elements.

\subsection{Elements and interactions}

The physical model has 4 types of elements, $A_i$, $A'_i$, $B_i$ and $C_i$, each existing in $N$ different types $i$. Interactions occur only between elements of the same type $i$, except for a few binding interactions between some $A_i$ and $C_j$ with $j\neq i$. These interactions are reciprocal and are represented as $i\sim j$. An element $A_i$ has two interacting interfaces, one through which it interacts with $A'_i$, $B_i$ and $C_j$ with $j\sim i$, and one through which it interacts with $C_i$. Elements $A_i'$ and $B_i$ have a single interface each and interact only with $A_i$. Elements $C_i$ have three interfaces through which they interact with $A_i$, $A_j$ with $j\sim i$, and $A'_i$. For simplicity, we assume that all elements of the same type have equivalent physical properties.

We denote by $h^-$ the activation energy for the dissociation of $A_iA'_i$ into $A_i+A'_i$, and $h^+$ the activation energy for the association of $A_i+A'_i$ into $A_iA'_i$. These activation energies are represented as
\beq\label{eq:AiAi}
A_i+A'_i\harp[h^-]{h^+}A_iA'_i,
\eeq
while the geometrical constraints are represented as 
\begin{figure}[H]
\centering
\includegraphics[width=.25\linewidth]{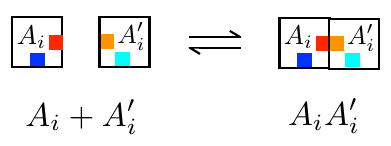}
\end{figure}

Similarly, the interaction of $A_i$ with $B_i$ is described as
\beq
B_i+A_i\harp[\phi_B^-]{\phi_B^+}B_iA_i
\eeq
with an activation energy $\phi_B^+$ for association and an activation energy $\phi_B^-$ for dissociation, and represented as
\begin{figure}[H]
\centering
\includegraphics[width=.2\linewidth]{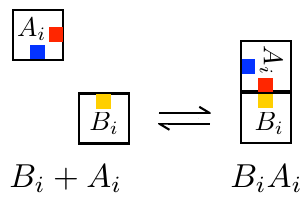}
\end{figure}

The interaction of $A_i$ with $C_i$ is described as
\beq\label{eq:CiAi}
C_i+A_i\harp[\e]{0}C_i\.A_i
\eeq
with no activation energy for association and an activation energy $\e$ for dissociation, and represented as
\begin{figure}[H]
\centering
\includegraphics[width=.2\linewidth]{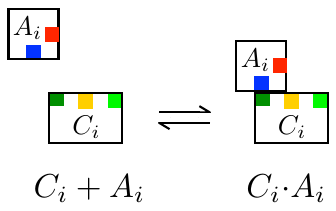}
\end{figure}
Note that $A_i$ interacts with $C_i$ through a different interface than with $A_i'$ or $B_i$. The difference is represented symbolically by denoting $C_i\.A_i$ with a dot.

The interaction of $A_i$ with $C_j$ with $j\sim i$ is described as
\beq
C_j+A_i\harp[\phi_C^-]{\phi_C^+}C_jA_i
\eeq
with an activation energy $\phi_C^+$ for association and an activation energy $\phi_C^-$ for dissociation, and represented as
\begin{figure}[H]
\centering
\includegraphics[width=.2\linewidth]{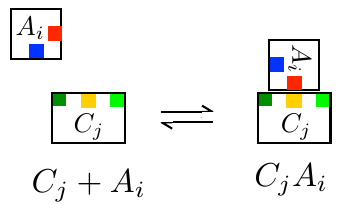}
\end{figure}
Note that $A_i$ interacts with $C_j$ through the same interface as with $A_i'$ and $B_i$. This ensures that $A_i$ can interact with $C_j$ but $A_iA'_i$ cannot.

Finally, the interaction of $A'_i$ with $C_i$ is described as 
\beq
C_i+A'_i\harp[\e]{0}C_i\.A'_i
\eeq
with the same activation energy $\e$ as in Eq.~\eqref{eq:CiAi}, and represented as
\begin{figure}[H]
\centering
\includegraphics[width=.2\linewidth]{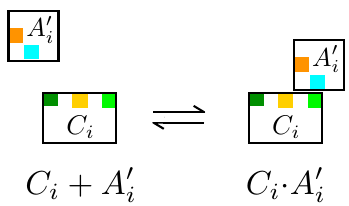}
\end{figure}

We ignore complexes $A_iB_iC_i$ (which could be excluded as a consequence of slightly modified geometries or by introducing  repulsion between $B_i$ and $C_i$).

\subsection{Detailed spontaneous reaction}

To account for catalysis, a more detailed description of the reaction $A_iA'_i\harp[h^-]{h^+}A_i+A'_i$ is needed that includes the transition state $(A_iA_i')^\ddagger$ when $A_i$ and $A_i'$ are at an intermediate distance of each other at which the potential energy is maximal,
\begin{figure}[H]
\centering
\includegraphics[width=.45\linewidth]{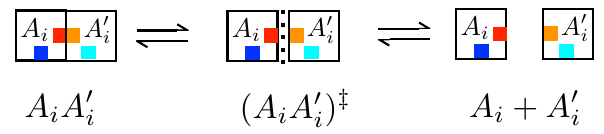}
\end{figure}
This corresponds to the following sequence of events where the energy of each state is indicated below each symbol:
\beq
\underset{0}{A_i+A'_i}\harp[]{}\underset{h^+}{(A_iA'_i)^\ts}\harp[]{}\underset{h^+-h-}{A_iA'_i}
\eeq
Taking differences of energies between successive states to define activation barriers, this is represented as
\beq
A_i+A'_i\harp[0]{h^+}(A_iA'_i)^\ts\harp[h^-]{0}A_iA'_i
\eeq
$(A_iA'_i)^\ts$ is an unstable intermediate and eliminating it leads to Eq.~\eqref{eq:AiAi}.

\subsection{Catalysis of bond cleavage}

The geometry of $C_i$ is chosen to enable the catalysis of the spontaneous reaction $A_iA'_i\harp[h^+]{h^-}A_i+A'_i$. The mechanism is represented by
\begin{figure}[H]
\centering
\includegraphics[width=.8\linewidth]{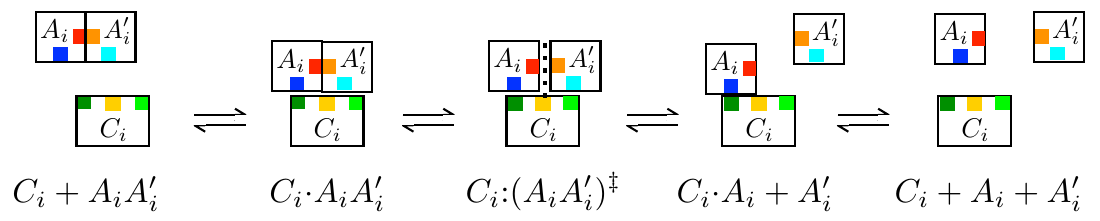}
\end{figure}
corresponding to a catalytic cycle described by
\beq
\underset{h^+-h^-}{C_i+A_iA'_i}\harp[]{}\underset{h^+-h^--\e}{C_i\.A_iA'_i}\harp[]{}\underset{h^+-2\e}{C_i{:}(A_iA'_i)^\ts}\harp[]{}\underset{-\e}{C_i\.A_i+A'_i}\harp[]{}\underset{0}{C_i+A_i+A_i}
\eeq
where an alternative path is for $A_i$ to be released before $A'_i$. This design is such that the two products $A_i$ and $A'_i$ cannot be fully attached to $C_i$, i.e., only in the transition state does $C_i$ has two interacting interfaces. This is achieved by proper positioning of the interacting interfaces~\cite{Munoz-Basagoiti.2023}.\\

Assuming $0<\e<h^\pm$, the cycle is equivalently described by activation energies as
\beq
C_i+A_iA'_i\harp[\e]{0}C_i\.A_iA'_i\harp[0]{h^--\e}C_i{:}(A_iA'_i)^\ts\harp[h^+-\e]{0}C_i\.A_i+A'_i\harp[0]{\e}C_i+A_i+A'_i
\eeq
Eliminating the unstable intermediate, we obtain
\beq\label{eq:catbc}
C_i+A_iA'_i\harp[\e]{0}C_i\.A_iA'_i\harp[h^+-\e]{h^--\e}C_i\.A_i+A'_i\harp[0]{\e}C_i+A_i+A'_i
\eeq
In this mechanism, catalysis is achieved by replacing a single barrier $h^-$ by two smaller barriers $h^--\e$ and $\e$. The catalytic rate is controlled by the largest of these two barriers, corresponding to an effective barrier $\max(h^--\e,\e)$ which is minimal for $\e=h^-/2$ (a more detailed analysis shows that this reasoning is valid only if $h^+>h^-$, otherwise the effective barrier is larger because of frequent recrossing events)~\cite{Rivoire.2023}. In what follows we take $h^+=\infty$ and simply denote $h^-$ by $h$, so that the requirement for catalysis reads
\beq
0<\e<h
\eeq

\subsection{All elementary reactions}

We consider a chemostat where some species are injected and all species are diluted at a constant rate $\d$. In the well-mixed case (where the diffusion constant $D$ is infinite), the model is described by the following reactions where, as above, we indicate above or below the arrows the reaction rate constants $k$ as $-(\ln k)/k_BT$ with $k_BT=1$ to fix the unit of energy, where $T$ is the temperature and $k_B$ the Boltzmann constant.

Injection:
\begin{align}\label{eq:inj}
\emptyset&\xa{-\ln(\d[R])}A_iA'_i\\
\emptyset&\xa{-\ln(\d[B]_0)}B_i\\
\emptyset&\xa{-\ln(\d[C]_0)}C_i
\end{align}
where $[R]$, $[B]_0$ and $[C]_0$ can be interpreted as the concentrations at which the molecules are injected with rate $\d$.

Spontaneous reaction:
\begin{align}\label{eq:spo}
&A_iA'_i\xa{h}A_i+A'_i
\end{align}

Catalysis:
\begin{align}
&C_i+A_iA'_i\harp[\e]{0} C_i\.A_iA'_i\label{eq:cat1}\\
&C_i\.A_iA'_i\xa{h-\e} C_i\.A_i+A'_i\\
&C_i\.A_iA'_i\xa{h-\e} C_i\.A'_i+A_i\\
&C_i\.A_i\harp[0]{\e} C_i+A_i\\
&C_i\.A'_i\harp[0]{\e} C_i+A'_i \label{eq:cat2}
\end{align}

Inhibition:
\begin{align}
B_i+A_i&\harp[\phi_B^-]{\phi_B^+}B_iA_i\\
C_j+A_i&\harp[\phi_C^-]{\phi_C^+}C_jA_i\qquad (j\sim i)
\end{align}

Dilution:
\begin{align}
A_i&\xa{-\ln\delta}\emptyset\\
A'_i&\xa{-\ln\delta}\emptyset\\
A_iA'_i&\xa{-\ln\delta}\emptyset\\
B_i&\xa{-\ln\delta}\emptyset\\
C_i&\xa{-\ln\delta}\emptyset\\
B_iA_i&\xa{-\ln\delta}\emptyset\\
C_i\.A_iA'_i&\xa{-\ln\delta}\emptyset\\
C_i\.A_i&\xa{-\ln\delta}\emptyset\\
C_i\.A'_i&\xa{-\ln\delta}\emptyset\\
C_jA_i&\xa{-\ln\delta}\emptyset\
\label{eq:dilAA}
\end{align}

Note that we do not distinguish between $C_i\.A_iA'_i$ and $C_i\.A'_iA_i$. These reactions are translated into ordinary differential equations by associating a kinetic rate with each activation barrier via the Arrhenius equation and ignoring differences in prefactors. For instance, the reaction $B_i+A_i\harp[\phi_B^-]{\phi_B^+}B_iA_i$ has a forward first-order kinetic rate $k_+=\kappa_0e^{-\phi_B^+}$ and backward second-order kinetic rate $k_-=\kappa_0\Omega_0 e^{-\phi_B^-}$ where $\kappa_0^{-1}$ represents an unit of time and $\Omega_0$ a unit of volume (the unit of energy is $k_BT=1$). We take $\kappa_0=1$ to fix the unit of time, while leaving $\Omega_0$ as a parameter. With these conventions, the equations describing a large well-mixed system are
\begin{align}
\frac{d [A_iA'_i]}{dt}&=\d[R]-e^{-h}[A_iA'_i]+e^{-\e}[C_i\.A_iA'_i]-\Omega_0[C_i][A_iA'_i]-\d[A_iA'_i]\\
\frac{d [A_i]}{dt}&=e^{-h}[A_iA'_i]+e^{-\e}[C_i\.A_i]-\Omega_0[C_i][A_i]+e^{-h+\e}[C_i\.A_iA'_i]+e^{-\phi_B^-}[B_iA_i]-e^{-\phi_B^+}\Omega_0[B_i][A_i]\nonumber\\
&+\sum_{j\sim i}(e^{-\phi_C^-}[C_jA_i]-e^{-\phi_C^+}\Omega_0[C_j][A_i])-\d [A_i]\\
\frac{d [A'_i]}{dt}&=e^{-h}[A_iA'_i]+e^{-h+\e}[C_i\.A_iA'_i]-\d[A'_i]+e^{-\e}[C_i\.A'_i]-\Omega_0[C_i][A'_i]\\
\frac{d [B_i]}{dt}&= \d[B]_0+e^{-\phi_B^-}[B_iA_i]-e^{-\phi_B^+}\Omega_0[B_i][A_i]-\d[B_i]\\
\frac{d [C_i]}{dt}&= \d[C]_0-\Omega_0[C_i][A_iA'_i]+e^{-\e}[C_i\.A_iA'_i]+e^{-\e}[C_i\.A_i]-\Omega_0[C_i][A_i]+e^{-\e}[C_i\.A'_i]-\Omega_0[C_i][A'_i]\nonumber\\
&+\sum_j(e^{-\phi_C^-}[C_iA_j]-e^{-\phi_C^+}\Omega_0[C_i][A_j])-\d [C_i]\\
\frac{d [B_iA_i]}{dt}&=e^{-\phi_B^+}\Omega_0[B_i][A_i]-e^{-\phi_B^-}[B_iA_i]-\d [B_iA_i]\\
\frac{d [C_i\.A_iA'_i]}{dt}&=\Omega_0[C_i][A_iA'_i]-e^{-\e}[C_iA_iA'_i]-2e^{-h+\e}[C_i\.A_iA'_i]-\d [C_i\.A_iA_i']\\
\frac{d [C_i\.A_i]}{dt}&=\Omega_0[C_i][A_i]-e^{-\e}[C_i\.A_i]+e^{-h+\e}[C_i\.A_iA'_i]-\d [C_i\.A_i]\\
\frac{d [C_i\.A'_i]}{dt}&=\Omega_0[C_i][A'_i]-e^{-\e}[C_i\.A'_i]+e^{-h+\e}[C_i\.A_iA'_i]-\d [C_i\.A'_i]\\
\frac{d [C_jA_i]}{dt}&=e^{-\phi_C^+}\Omega_0[C_j][A_i]-e^{-\phi_C^-}[C_jA_i]-\d [C_jA_i]
\end{align}

\section{Coarse-graining}\label{app:cg}

The set of reactions is described by a smaller number of effective reactions when the different reaction rates are separated in time.

\subsection{Microscopic model}

 To perform the reduction, we rewrite all the reactions from slowest to fastest. We indicate above and below the arrows the reaction rate constants (rather than their activation barrier as above). The indices refer to the time scale of these rate constants.
\begin{align}
&A_iA_i'\xa{k_{-1}}A_i+A'_i &(\tau\sim \ve^{-1})\\
&\emptyset \xa{\d[R]_0}A_iA_i',\qquad \emptyset\xa{\d[C]_0}C_i &(\tau\sim 1)\\
&A_iA_i',\ C_i, \ A_i, \ A'_i,\ C_i\.A_iA'_i,\ C_i\.A_i,\ C_i\.A'_i,\ C_j\-A_i\xa{\d}\emptyset &(\tau\sim 1)\\
&C_i\.A_iA'_i\xa{k_0}C_i\.A_i+A'_i,\qquad C_i\.A_iA'_i\xa{k_0}C_i\.A'_i+A_i&(\tau\sim 1)\\
&A_i+C_j\harp[k_1^-]{k_1^+}C_j\-A_i&(\tau\sim \ve)\\
&A_i+B_i\harp[k_2^-]{k_2^+}B_i\-A_i &(\tau\sim \ve^2)\\
&A_i+C_i\harp[k_3^-]{k_3^+}C_i\.A_i,\qquad A'_i+C_i\harp[k_3^-]{k_3^+}C_i\.A'_i,\qquad A_iA_i'+C_i\harp[k_3^-]{k_3^+}C_i\.A_iA'_i &(\tau\sim \ve^3)
\end{align}
The time scale of each of these reactions is given in parentheses on the right, with a parameter $\ve$ that must be send to zero for the coarse-graining to be mathematically exact~\cite{Bo.2017} (this adimensional $\ve$ is distinct from the $\e$ that denotes an activation energy). The correspondence to the previous notations is as follows
\beq
k_{-1}=e^{-h},\quad k_0=e^{-(h-\e)},\quad k_1^+=\Omega_0 e^{-\phi_C^+},\quad k_1^-=e^{-\phi_C^-},\quad k_2^+=\Omega_0e^{-\phi_B^+},\quad k_2^-=e^{-\phi_B^-},\quad k_3^+=\Omega_0,\quad k_3^-=e^{-\e}
\eeq

\subsection{Principle of the reduction}

We consider the following hierarchy of timescales
\beq
k_{-1}\ll k_0,\d\ll k_1^\pm\ll k_2^\pm\ll k_3^\pm
\eeq
Formally, this is controlled by the small quantity $\ve$ by considering $k_n^\pm=\ve^{-n}\bar k_n^\pm$ where $\bar k_n^\pm$ are of order 1.

In addition, to simplify the formula, we also need $k_n^+/k_n^-$ to be either large or small: an assumption of strong or weak binding that corresponds to a separation of timescale within the previous one. This is controlled by an other small quantity $\ve'$ larger than $\ve$: $\ve\ll\ve'\ll 1$. Formally, we take the limits $\ve\to 0$ and $\ve'\to 0$ with $\ve/\ve'\to 0$. This is achieved, for example, when $\ve'=\ve^{1/2}$ with $\ve\to 0$.

Overall, we assume
\beq\label{eq:hierarchy}
k_{-1}\ll k_0,\d\ll k_1^+\ll k_1^-\ll k_2^-\ll k_2^+\ll k_3^+\ll k_3^-
\eeq
The constraints $k_{-1}\ll k_0\ll k_3^-\ll k_3^+$ impose
\begin{align}
&0<\e<h/2\\
&\e<-\ln\Omega_0
\end{align}
The other kinetic rates each involve an independent parameter which ensures that Eq.~\eqref{eq:hierarchy} can be satisfied.

\subsection{Elimination of processes on the slowest timescale $\tau\sim\ve^{-1}$}

We ignore the spontaneous reaction which occurs on a timescale much slower than the other ones.

\subsection{Elimination of processes on the fastest timescale $\tau\sim \ve^3$}

On the fastest timescale, we have a quasi-equilibrium of the reactions
\beq
A_i+C_i\harp[k_3^-]{k_3^+}C_i\.A_i, \qquad A'_i+C_i\harp[k_3^-]{k_3^+}C_i\.A'_i,\qquad A_iA_i'+C_i\harp[k_3^-]{k_3^+}C_i\.A_iA'_i 
\eeq

The slow variables are 
\begin{align}
[A_i]_3&=[A_i]+[C_i\.A_i]\\ 
[A'_i]_3&=[A'_i]+[C_i\.A'_i]\\ 
[A_iA'_i]_3&=[A_iA'_i]+[C_i\.A_iA'_i]\\ 
[C_i]_3&=[C_i]+[C_i\.A_i]+[C_i\.A'_i]+[C_i\.A_iA'_i] 
\end{align}
They are defined to be unchanged by the fast processes. Assuming equilibrium on the fast timescale, we obtain $[C_i]$, $[A_i]$ and $[C_iA_iA'_i]$ as function of $[A_i]_3$, $[C_i]_3$ etc. The slow variables are simply expressed in terms of the fast variables, for instance
\begin{align}
[A_i]_3&=[A_i](1+k_3^+/k_3^-[C_i])\\ 
[C_i]_3&=[C_i](1+k_3^+/k_3^-([A_i]+[A'_i]+[A_iA'_i]))
\end{align}
but inverting these relationships may not lead to simple expressions.

On the slower timescale, the effective dynamics is given by
\begin{align}
\p_t[C_i]_3&=\d([C]_0-[C_i]_3)+\sum_{j\sim i}(k_1^-[C_i\-A_j]-k_1^+[C_i][A_j])\\
\p_t[A_i]_3&=k_0[C_i\.A_iA'_i]-\d[A_i]_3-\sum_{j\sim i}(k_1^-[C_j\-A_i]-k_1^+[C_j][A_i])-(k_2^-[B_i\-A_i]-k_2^+[A_i][B_i])\\
\p_t[A_iA_i']_3&=\d([R]_0-[A_iA'_i]_3)-2k_0[C_i\.A_iA'_i]
\end{align}
where $[C_i]$, $[A_i]$ and $[C_i\.A_iA'_i]$ are understood as functions of $[A_i]_3$, $[A_iA'_i]_3$ and $[C_i]_3$ so that we have effectively eliminated them.

To simplify the formula, we take the limit $k_3^+/k_3^-\ll 1$ where $[A_i]_3\simeq [A_i]$, $[A_iA'_i]_3\simeq [A_iA'_i]$, $[C_i]_3\simeq [C_i]$ and $[C_iA_iA'_i]\simeq k_3^+/k_3^-[C_i][A_iA'_i]$, leading to
\begin{align}
\p_t[C_i]&=\d([C]_0-[C_i])+\sum_{j\sim i}(k_1^-[C_i\-A_j]-k_1^+[C_i][A_j])\\
\p_t[A_i]&=k_0k_3^+/k_3^-[C_i][A_iA'_i]-\d[A_i]-\sum_{j\sim i}(k_1^-[C_j\-A_i]-k_1^+[C_j][A_i])-(k_2^-[B_i\-A_i]-k_2^+[A_i][B_i])\\
\p_t[A_iA_i']&=\d([R]_0-[A_iA'_i])-2k_0k_3^+/k_3^-[C_i][A_iA'_i]
\end{align}
More precisely, $k_3^+/k_3^-\sim\ve'\gg\ve$ so this limit preserves the (other) timescale separation.
 
\subsection{Elimination of processes on the second fastest timescale $\tau\sim \ve^2$}

On the second fastest timescale, we have a quasi-equilibrium of the reaction
\beq
A_i+B_i\harp[k_2^-]{k_2^+}B_i\-A_i\\
\eeq

The slow variables are  
\begin{align}
[A_i]_2&=[A_i]+[B_i\-A_i]\\
[B_i]_2&=[B_i]+[B_i\-A_i]
\end{align}
in term of which we can express the others using the equilibrium constants. We have an effective dynamics given by
\begin{align}
\p_t[A_i]_2&=k_0k_3^+/k_3^-[C_i][A_iA'_i]-\d[A_i]_2-\sum_{j\sim i}(k_1^-[C_j\-A_i]-k_1^+[C_j][A_i])\\
\p_t[B_i]_2&=\d([B]_0-[B_i]_2)
\end{align}
where $[A_i]$ is understood as function of the slower variables $[A_i]_2$ and $[B_i]_2$  so that we effectively eliminated $[A_i]$, $[B_i]$ and $[B_i\-A_i]$. The equations for $[C_i]$ and $[A_iA'_i]$ are unchanged.

A first simplification is to replace $[B_i]_2$ by $[B]_0$, which is justified since if $[B_i]_2=[B]_0$ at any time, it remains constant.

A second simplification is to remark that the equation for $[A_i]$ as a function of $[A_i]_2$ and $[B_i]_2$ simplifies in the limit $k_2^-/k_2^+\ll 1$ to become
\beq
[A_i]\simeq\max(0,[A_i]_2-[B]_0)
\eeq
This equation has a simple interpretation: in the limit of strong affinity, $B_i$ is saturated by any available $A_i$.
More precisely, we consider $k_2^-/ k_2^+=\ve'$; note that it is a limit of strong binding, opposite to the limit of weak binding made with $k_3^+/ k_3^-=\ve'$.

\subsection{Elimination of processes on the third fastest timescale $\tau\sim \ve$}

On the third fastest timescale, we have a quasi-equilibrium of the reaction
\beq
A_i+C_j\harp[k_1^-]{k_1^+}C_j\-A_i\qquad (j\sim i)\
\eeq
 
The slow variables are  
\begin{align}
[A_i]_1&=[A_i]+\sum_{j\sim i}[C_j\-A_i]\\
[C_i]_1&=[C_i]+\sum_{j\sim i}[C_i\-A_j]
\end{align}
in term of which we can express the others. We have an effective dynamics given by
\begin{align}
\p_t[C_i]_1&=\d([C]_0-[C_i]_1)\\
\p_t[A_i]_1&=k_0k_3^+/k_3^-[C_i][A_iA'_i]-\d[A_i]_1
\end{align}
where $[C_i]$ is understood as a function of the slow variables $[A_i]_1$ and $[C_i]_1$. 

We have 
\beq
[C_i]=\frac{[C_i]_1}{1+k_1^+/k_1^-\sum_{j\sim i}[A_j]}\simeq \frac{[C_i]_1}{1+k_1^+/k_1^-\sum_{j\sim i}\max(0,[A_j]_2-[B]_0)}
\eeq

A first simplification is to assume $[A_i]_1\simeq [A_i]_2$ which is achieved provided $k_1^+/k_1^-=\G\ve'$ where $\G$ is of order 1 (limit of weak binding). 

A second simplification is to assume $[C_i]_1\simeq [C]_0$ which is justified since $[C_i]_1= [C]_0$ remains constant.

A third simplification is to remark that $k_0k_3^+/k_3^-\ll \d$ since we assumed $k_3^+/k_3^-=\ve'$ so that
\beq
\p_t[A_iA_i']=\d([R]_0-[A_iA'_i])-2k_0k_3^+/k_3^-[C_i][A_iA'_i]
\eeq
can be approximated by
\beq
\p_t[A_iA_i']=\d([R]_0-[A_iA'_i])
\eeq
and we can consider $[A_iA'_i]\simeq [R]_0$.

\subsection{Coarse-grained model}

The final result is 
\beq
\p_t[A_i]_1=\frac{k_3^+/k_3^-k_0[C]_0[R]_0}{1+k_1^+/k_1^-\sum_{j\sim i}\max(0,[A_j]_1-[B]_0)}-\d [A_i]_1
\eeq

Or in terms of order 1 quantities only, since $k_3^+/k_3^-=\ve'$ and $k_1^+/k_1^-=\ve'\G$,
\beq
\p_t[A_i]_1=\frac{\ve'k_0[C]_0[R]_0}{1+\ve'\G\sum_{j\sim i}\max(0,[A_j]_1-[B]_0)}-\d [A_i]_1
\eeq

We can eliminate $\ve'$ from this formula by rescaling the volume so that $[A_i]'_1=\ve'[A_i]_1$, $[B]'_0=\ve'[B]_0$, $[C]'_0=\ve'[C]$, $[R]'_0=\ve'[R]_0$.
This leads to
\beq
\p_t[A_i]'_1=\frac{k_0[C]'_0[R]'_0}{1+\G\sum_{j\sim i}\max(0,[A_j]'_1-[B]'_0)}-\d [A_i]'_1
\eeq
Denoting $\L=k_0[C]'_0[R]'_0$, we arrive at Eq.~(1).
The remaining parameters are free to take any value of order 1 relative to $\ve'$ and $\ve$.

The coarse-grained model corresponds to two effective reactions,
\begin{align}
\emptyset&\xa{k_i}A_i\label{eq:cg1}\\
A_i&\xa{\delta}\emptyset\label{eq:cg2}
\end{align}
with an effective rate $k_i$ of the form
\beq
k_i=\frac{\Lambda}{1+\G\sum_{j\sim i}\max(0,[A_j]-B)}
\eeq

\section{Demographic noise}\label{app:demo}

We add to the deterministic description fluctuations arising from the finite number of molecules, leading to Poissonian noise in the actual realized number of reactions of each type.
We follow standard and general procedures~\cite{gardiner1985handbook} that can be applied to both the microscopic and coarse-grained models, with or without diffusion.

\subsection{Reaction rates and the Master equation}

Without loss of generality, any system of reactions involving $S$ species $i$ can be decomposed into a set of unidirectional reactions of the form
\beq
\sum_i\nu_{\ell i}^-X_i\xa{k_\ell(N)}\sum_i\nu_{\ell i}^+X_i
\eeq
where $X_i$ is a symbol for species $i$, $\nu_{\ell i}^\pm$ are stoichiometric coefficients and $k_\ell(N)$ is a reaction rate constant. Here $N$ is an $S$-dimensional vector representing the molecular number of each species $i$ and $\ell$ indexes the reaction. This is associated with a master equation of the form
\beq\label{eq:master}
\p_t P(N)=\sum_\ell [a_\ell(N-r_\ell)P(N-r_\ell)-a_\ell(N)P(N)]
\eeq
with, assuming mass action kinetics, 
\begin{align}
&r_\ell=\sum_i(\nu_{\ell i}^+-\nu_{\ell i}^-)e_i\\
&a_\ell(N)=k_\ell(N)\Omega\prod_i \frac{N_i!}{(N_i-\nu_{\ell i}^-)!\Omega ^{\nu_{\ell i}^-}}
\end{align}
where $e_i$ is an $S$-dimensional vector with components $(e_i)_j=\delta_{ij}$ and where $\Omega$ is the volume. 

\subsection{Reduction using timescale separation}

In the deterministic coarse-graining procedure (Appendix~\ref{app:cg}),
we used timescale separation to reduce the equations to simpler ones.
A key result that we now show is that the noise of the fast processes can be neglected,
and that the noise of the remaining slowest processes corresponds
to the Poisson noise on the fluxes of the coarse-grained rate equations.

To show this, we follow the framework of \cite{lazarescu2019large} and take a timescale
$\delta t$ which is short compared to the rates, $k\delta t\ll1$,
so that the concentrations do not change much, but where the absolute number changes in this time are large, $k\Omega\delta t\gg1$ where $\Omega$ represents the volume. The
deterministic rates of the processes in Eqs.~(\ref{eq:inj}-\ref{eq:dilAA}) are given by $\Omega k_{\ell}(N)$ for process $\ell$. The number of events $\lambda_{\ell}(t)$ for
process $\ell$ between $t$ and $t+\delta t$ is Poisson-distributed
with mean $\Omega k_{\ell}(N)\delta t$, so that, for large $\Omega$, the probability $P(\lambda_{\ell}(t))$ satisfies
\beq
\ln P(\lambda_{\ell}(t)) \simeq -\Omega k_{\ell}(N)\delta t \phi\left(\frac{\lambda_{\ell}(t)}{\Omega k_{\ell}(N)\delta t}\right)
\eeq
where $\phi(x)\equiv x\ln x-x+1$ has a single
minimum at $x=1$. For $\delta t$ chosen as prescribed, the $\lambda_{\ell}(t)$
are independent~\cite{lazarescu2019large}, so the probability of all rates is
simply
\beq
\ln P[\{\lambda_\ell\}_\ell]=
\sum_{\ell}\ln P(\lambda_{\ell}(t))=-\Omega\delta t\sum_{l}k_{\ell}(N)\phi\left(\frac{\lambda_{\ell}(t)}{\Omega k_{\ell}(N)\delta t}\right)
\eeq

We now consider a situation where there is timescale separation, so that the rates $k_{\ell'}$
of the fast processes $\ell'\in \mathcal{F}$ are larger by a factor of $1/\epsilon$ compared to others $\ell''\in \mathcal{S}$ ($k_{\ell'}=\tilde k_{\ell'}/\epsilon$ with $\tilde k_{\ell'}\sim k_{\ell''}$),
\beq
\ln P[\{\lambda_\ell\}_\ell]=-\Omega\delta t\left[\frac{1}{\epsilon}\sum_{\ell'\in\mathcal{F}}\tilde k_{\ell'}(N)\phi\left(\frac{\lambda_{\ell'}(t)}{\Omega k_{\ell'}(N)\delta t}\right)+\sum_{\ell''\in\mathcal{S}}k_{\ell''}(N)\phi\left(\frac{\lambda_{\ell''}(t)}{\Omega k_{\ell''}(N)\delta t}\right)\right]
\eeq
Then, deviations of $x=\lambda_{\ell'}(t)/(\Omega k_{\ell'}(N)\delta t)$
from $x=1$, where $\phi(x)=0$, are small, since they are highly unlikely,
with contributions scaling with $\epsilon$ as, $P\sim e^{-\kappa\epsilon^{-1}}$. We can therefore assume that the fast processes happen at the deterministic rate given by $\lambda_{\ell'}(t)=\Omega k_{\ell'}(N)\delta t$. More precisely, this assumes that the fluctuations of the fast processes 
are at most of the order of the typical contribution of the slow terms: since $k_{\ell}=O(1/\epsilon)$, we allow $\phi\left(\frac{\lambda_{\ell'}(t)}{\Omega k_{\ell'}(N)\delta t}\right)=O(\epsilon)$, i.e., $\frac{\lambda_{i}(t)}{\Omega k_{\ell'}(N)\delta t}=1+O(\sqrt{\epsilon})$. From the deterministic
derivation of Appendix~\ref{app:cg}, this implies that
the fast processes are kept at a quasi-equilibrium.

In summary, accounting for demographic noise in the coarse-grained model can be done at the level of the effective slow reactions given by Eqs.~\eqref{eq:cg1}-\eqref{eq:cg2} without considering the noise associated with the faster reactions that they encompass.

\subsection{Langevin equation for typical fluctuations}

For the purpose of numerical simulations, it is useful to make a further
approximation, by assuming that the fluctuations of $\lambda_{\ell}(t)$ deviates from its mean as the standard deviation of the distribution, $\left[\Omega k_{\ell}(N)\delta t\right]^{1/2}$.
In this case, the Poisson distribution can be approximated by a Gaussian,
giving the chemical Langevin equation~\cite{Schnoerr.2017} where
\beq\label{eq:generalCL}
\frac{dN_i}{dt}=\sum_\ell\left[r_{\ell i} a_\ell(N)+ r_{\ell i}\sqrt{a_\ell(N)}\xi_\ell(t)\right] 
\eeq
with $\langle\xi_\ell(t)\xi_{\ell'}(t)\rangle=\delta_{\ell\ell'}\delta(t-t')$, in the It\^o convention, corresponding to an independent contribution to the noise from each chemical reaction.

We verify that this approximation is justified for the range of parameters that we consider. In particular, flip events occur when one active variable decreases to allow for another to increase, as argued in Appendix~\ref{app:wm}. These events happen when an active variable $[A_1]$ spontaneously decreases to about $B+O(1/\Gamma)$ (and in fact even at higher $[A_1]$, since for finite $\Gamma$ the inactive variable also increases at the same time). In the relevant range, $[A_1]\gtrsim B+O(1/\Gamma)$ and the distribution of $N_1=\Omega[A_1]$ shows almost no difference between the two noises, as shown in Fig.~\ref{fig:Gauss_vs_Poiss}. Since sampling Poisson random variables is computationally much more expensive than sampling Gaussian ones, we use Gaussian noise in our simulations.

\subsection{The chemical Langevin equation for the well-mixed coarse-grained model}

At the coarse-grained level, the fluctuations of all constituents but $A_i$ can be neglected, as argued above. We have only two effective reactions
\begin{align}
C_i\empty&\xa{}C_i+A_i\\
A_i&\xa{}\emptyset
\end{align}
Let $N$ be the vector whose components $N_i$ are the number of $A_i$ and $C_i(N)$ be defined by
\beq
C_i(N)=\frac{\Lambda\Omega}{1+\G\sum_{j\sim i}\max(0,N_j/\Omega-[B_j])}
\eeq
where $\Omega$ is the volume.  
The master equation has the form
\beq
\partial_tP_t(N)=\sum_i C_i(N-e_i)P_t(N-e_i)+\d(N+e_i)P_t(N+e_i)-(C_i(N)-\d N)P_t(N)
\eeq
and the chemical Langevin equation is therefore
\beq
\partial_tN_i=C_i(N)-\d N_i+\sqrt{C_i(N)}\ \xi^{(c)}_{i}(t)+\sqrt{\d N_i}\ \xi^{(d)}_{i}(t)
\eeq
which simplifies to
\beq
\partial_tN_i=C_i(N)-\d N_i+\sqrt{C_i(N)+\d N_i}\ \xi_i(t)
\eeq
where $\xi^{(c)}_{i}(t)$, $\xi^{(d)}_{i}(t)$ and $\xi_i(t)$ are Gaussian white noises with unit variance.

With $n_i=N_i/\Omega$ and $c_i(n)=C_i(N=\Omega n)/\Omega$, i.e.,
\beq
c_i(n)=\frac{\Lambda}{1+\G\sum_{j\sim i}\max(0,n_j-[B_j])}
\eeq
we obtain
\beq\label{eq:chemL}
\partial_tn_i=c_i(n)-\d n_i+\omega \sqrt{c_i(n)+\d n_i}\ \xi_i(t)
\eeq
where
\beq
\omega=\frac{1}{\sqrt{\Omega}}
\eeq
By redefining time as $t'=\d t$ and concentrations as $[A_i]'=\d[A_i]/\L$, we can consider $\G'=\L\G/\d$ and $\rho'=\rho\d/\L$ so as to obtain effectively $\L=1$ and $\d=1$. Only three parameters are left, $\G$, $[B_j]=[B]_0$ and $\omega$. 

\section{Model with diffusion}\label{app:D}

When space is discretized into cells of extension $\Delta x$, diffusion can be formally described by additional ``reactions'' describing the transfer of particles from one cell to the next. In the context of a one-dimensional system, the additional reactions for $A_i$ are
\begin{align}
&A_i(x)\xa{D/\D x^2}A_i(x+\D x)\\
&A_i(x)\xa{D/\D x^2}A_i(x-\D x)
\end{align}
where $x\pm\D x$ is understood modulo the total space length when considering periodic boundary conditions.

\subsection{Coarse-grained model with diffusion}

Since $A_i$ is involved in 4 diffusion reactions, each with its own noise, the extension of Eq.~\eqref{eq:chemL} to include one-dimensional diffusion takes the form
\begin{align}
&\partial_tn_i(x)=c_i(n(x))-\d n_i(x)+D/\D x^2 (n_i(x+\D x)+n_i(x-\D x)-2n_i(x))+\omega \sqrt{c_i(n)+\d n_i}\xi_i(x,t)\nonumber\\
&+\omega\sqrt{D}/\D x\left(\sqrt{n_i(x+\D x)}\xi_i^-(x+\D x)+\sqrt{n_i(x-\D x)}\xi_i^+(x-\D x)-\sqrt{n_i(x)}(\xi_i^-(x,t)+\xi_i^+(x,t))\right)
\end{align}
In the continuous limit $\Delta x\to 0$, it is more concisely written
\beq
\partial_tn_i(x)=c_i(n(x))-\d n_i(x)+D\partial_x^2n_i(x)+\omega \sqrt{c_i(n)+\d n_i}\xi_i(x,t)+\omega\sqrt{2D}\partial_x\left(\sqrt{n_i(x)}\zeta_i(x,t)\right)
\eeq
where $\xi_i(x,t)$ and $\zeta_i(x,t)$ are Gaussian white noises.
By rescaling space, we can always assume $D=1$.

\section{Steady-state distribution of well-mixed systems}\label{app:wm}

The results of numerical simulations reported in Fig.~2C suggest that the productivity increases on average with time. Here we explain the mechanism behind this behavior
for well-mixed systems. To do so, we calculate the transition probability
between states in a limiting case that can be analyzed analytically
and leads to a simple picture, the low-noise limit where
$\omega=\Omega^{-1/2}$ is small. This limit ensures that the ``active'' and ``inactive'' variables are well defined and that transitions are rare.
In this limit, we expect an Arrhenius-like behavior for the transition
rates, scaling as $\exp\left[-\frac{1}{\omega^{2}}g(\Gamma,B,\delta,\Lambda)\right]$
(this is called ``to exponential accuracy'' below.) We are
interested in the form of the function $g$ . In addition, we consider
the limit where $\Gamma$ is large and $B$ is small while keeping
$1/\Gamma\ll B\ll1$. This is in line with the requirement for bistability
of two variables, where $\Gamma$ needs to be large enough, and $1/\Gamma<B$.
After this limit is taken, we obtain $g(\delta,\Lambda)$ independent of $\Gamma$ and $B$.

The derivation below reveals the basic mechanism for
the growth in productivity: when $\Gamma$ is large, the inhibited
variables remain at low values, since it requires very strong noise for a variable to rise while it is inhibited. Transitions are therefore
more likely to occur due to a spontaneous decline of the values of
active variables below $B$, at which point they no longer inhibit
others. This makes transitions in which less active variables are inactivated more likely than their opposite counterparts, and leads to a preference for states with more active variables. This mechanism extends to models with unequal values of $\Lambda_i$.

Here we first derive the transition probability for one variable,
then for two variables, before generalizing to any number of variables.
From the transition probabilities, we obtain the stationary probability
of each state.

\subsection{System with a single variable}

For a single isolated variable, the slow effective reactions Eqs~\eqref{eq:cg1}-\eqref{eq:cg2} are simply
\begin{align}
\emptyset&\xa{\Lambda}A_1\\
A_1&\xa{\delta}\emptyset
\end{align}
Let $N$ be the number of $A_{1}$ particles.
$N\ge0$ satisfies a Master equation with rates $w_{N\to N+1}=\Lambda\Omega$
and $w_{N+1\to N}=\delta N$ whose stationary distribution is Poissonian,
\beq
P_{\text{alone}}(N)=\frac{\left(\Lambda\Omega/\delta\right)^{N}}{N!}e^{-\Lambda\Omega/\delta}\simeq\exp\left[-\frac{\Lambda\Omega}{\delta}\phi\left(\frac{\delta N}{\Lambda\Omega}\right)\right]\ .
\eeq
where the second equality is to exponential accuracy, with $\phi(x)\equiv x\ln x-x+1$. Since $\phi(x)$ has a single minimum at $x=1$, $P_{\text{alone}}(N)$
is maximal when $\delta N^{*}/(\Lambda\Omega)=1$, i.e. when $\left[A\right]^{*}=\Lambda/\delta$ as expected from the deterministic equations.

\subsection{System with two variables}

Now consider a system with two variables that reciprocally inhibit each other. If $N_{1}$ fluctuates around
the noiseless value $N_{1}^{*}=\Omega\Lambda/\delta$ (``active'' state),
then $N_{2}$ fluctuated around the deterministic value

\begin{equation}
N_{2}^{*}=\frac{\Omega\Lambda/\delta}{1+\Gamma\max(0,N_{1}^{*}/\Omega-B)}=\frac{\Omega\Lambda/\delta}{1+\Gamma(\Lambda/\delta-B)}
\end{equation}
where we assumed that $\Gamma>1/B$ and $\Lambda/\delta>B$.

How does a transition occur? When $\Gamma$ is large, it is very unlikely that the noise on $N_{2}$ will push $N_{2}$ up against the strong deterministic downward force. Therefore, a transition event at low noise typically requires
$N_{1}$ to go down first, until the downward force acting on $N_{2}$
no longer scales with $1/\Gamma$, namely when $\Gamma\max(0,N_{1}/\Omega-B)=\Gamma\left(N_{1}/\Omega-B\right)\lesssim1$.
If $N_{1}$ reaches $N_{1}<B\Omega$, this is enough, since then it has no inhibitory effect anymore. All in all, $N_{1}$ must go down to 
\begin{equation}
N_{1}^{\text{(flip)}}=\Omega(B+O(1/\Gamma))\ .
\end{equation}
What is the probability of $N_{1}$ going from $\Omega\Lambda/\delta$ to
$N_{1}^{\text{(flip)}}$? For small noise, $N_{1}$ spends most of
the time fluctuating close to its stable fixed point $N_{1}^{*}$.
The probability of an excursion from there at small noise, is to exponential
accuracy, the stationary probability $N_{1}^{\text{(flip)}}$
\beq
\ln P\left(N_{1}^{\text{(flip)}}\right)=-\frac{\Lambda\Omega}{\delta}\phi\left(\frac{\delta B}{\Lambda}+O(1/\Gamma)\right)\ .
\eeq
This can also be derived from a first-passage time calculation~\cite{doering2005extinction}.

After reaching $N_{1}^{\text{(flip)}}\simeq \Omega B$, $N_{1}$ must stay
at this value for long enough to allow for $N_{2}$ to grow above
$\Omega B$, at which point it will inhibit $N_{1}$, allowing the transition
to be completed. Since $N_{2}$ starts close to zero, and grows
as $\partial_{t}N_{2}=\Omega\Lambda-\delta N_{2}$, its evolution from this
point on is $N_{2}(t)=\Omega\Lambda/\delta(1-e^{-\delta t})$, and the
time it takes to reach $\Omega B$ is $t=\delta^{-1}\ln\frac{1}{1-B\delta/\Lambda}\simeq B/\Lambda+O\left(B^{2}\right)$.
The probability to remain below $N_{1}^{\text{(flip)}}$ decays exponentially
with time and so
contributes a term of the form $e^{-\Omega \kappa t}=e^{-\Omega \kappa B+O(B^{2})}$
where $\kappa$ is a constant. We now take $B$ to be small and
this term is negligible. All together, taking $B\to0$ (but keeping
$1/\Gamma\ll B$).
\begin{align}
\ln P_{\text{one transition}} & \sim-\Omega\left[\frac{\Lambda}{\delta}\phi\left(\frac{\delta}{\Lambda}B\right)+O(B)\right]\to-\Omega\frac{\Lambda}{\delta}=-\frac{\Lambda}{\omega^{2}\delta}
\end{align}
where we used $\phi(x\to0)=1$.

\subsection{Generalization and effective detailed-balance}

The above argument generalizes to any transition between states with
\begin{equation}
P_{\text{transition}}=P_{\text{one transition}}^{k}
\end{equation}
where $k$ is the number of variables that have to go from active to inactive.
From this it follows that the jump process satisfies a detailed balance
condition. Indeed, consider two states with $F_{1}$ and $F_{2}$ active variables, respectively, with $F_{\text{overlap}}$ active variables in common.
Then the transition $1\to2$ is $P_{\text{transition}}(1\to2)=P_{\text{one transition}}^{F_{1}-F_{\text{overlap}}}$
and similarly in the other direction. So 
\beq
\frac{P_{\text{transition}}(1\to2)}{P_{\text{transition}}(2\to1)}=\frac{P_{\text{one transition}}^{-F_{2}}}{P_{\text{one transition}}^{-F_{1}}}
\eeq
This means that for the coarse-grained process of jumps between stable
states satisfies detailed balance with an equilibrium distribution
\beq
P_{\text{eq}}=\left(P_{\text{one transition}}\right)^{-F}
\eeq
where here $F$ is the number of active variables in a state. This holds to the same level
of approximation as above (in particular, $\ln P_{\text{eq}}$ to
order $\omega^{-2}$), so that
\beq\label{eq:PF}
P_{\text{eq}}\simeq\exp\left(\frac{\Lambda F}{\omega^{2}\delta}\right)
\eeq
This equilibrium distribution corresponds to the frequency with which each stable state is sampled in the nonequilibrium steady state of our model.

\subsection{Different values of $\Lambda_{i}$}

When $\Lambda_{i}$ depends on $i$, the
above arguments generalize to lead to
\beq
P_{\text{eq}}\simeq\exp\left[\frac{1}{\omega^{2}\delta}\sum_{i\in\text{active}}\Lambda_{i}\right]
\eeq
where the productivity $\sum_{i\in {\rm active}}\Lambda_i$ reduces to $\Lambda F$ as in Eq.~\eqref{eq:PF} if all the $\Lambda_{i}$ are equal.

\section{Numerical methods}\label{app:num}

\subsection{Numerical implementation}

We simulate the chemical Langevin equation using the Euler-Maruyama algorithm, to which we add a maximum function to enforce positivity. The generic Langevin equation Eq.~\eqref{eq:generalCL} is thus discretized as
\beq
x_i(t+\delta t)=\max\left(0,x_i(t)+\sum_\ell\left[r_{\ell i} a_\ell(x)\delta t+ r_{\ell i}\sqrt{a_\ell(x)}w_\ell(t)(\delta t)^{1/2}\right]\right),\qquad w_\ell(t)\sim\N(0,1)
\eeq
We take $\delta t=10^{-2}$ in a context where other parameters are of order 1.

We discretize space by considering a one-dimensional system with periodic conditions divided into $M$ cells with $D=1$ and run simulations for a given total time $T$. We record $K<T/\Delta t$ time points distributed in $[0,T]$ on a logarithmic scale. The output of a simulation is therefore a $K\times M\times N$ array $X$ of concentrations $X_{kmn}\geq 0$.

\subsection{Visual representation}

To represent time evolution and spatial diffusion, we collapse $X_{kmn}$ into the $K\times M$ array $Y_{km}$ and represent the third dimension by a color code. Following approaches developed in analytical chemistry~\cite{gardner2020self}, we use a coloring scheme based on similarity between compositions. To do this, we first perform a low-dimensional projection of the trajectories by  UMAP (Uniform Mapping Approximation and Projection)~\cite{mcinnes2018umap}. Specifically, we first transform the $K\times M\times N$ output into a $KM\times N$ matrix and then apply an implementation of UMAP in Julia, UMAP.jl~\cite{UMAP_jl}, using 3 components, the metric \texttt{corr\_dist}, and otherwise default parameters. The result is a $KM\times 3$ output that we map to the CIELAB color space, which is designed so that numerical differences between values correspond to the amount of change humans see between colors~\cite{CIE2007CIELAB}.

\subsection{\label{app:states}States and productivity}

Stable states are defined as combinations of active and inactive $A_i$, represented by 1 and 0, respectively. A stable state formally corresponds to an independent set of the graph of inhibitory interactions, where $A_i$ is active if and only if all $A_j$ to which it is connected by inhibitory interactions are suppressed. All stable states can be enumerated using the Bron-Kerbosch algorithm~\cite{bron1973algorithm,eppstein2010listing}, which is used for panels C and D of Fig.~3.

There are two ways to assign a state to a given composition $(X_{km1},\dots,X_{kmN})$. One is to note that the distribution of concentrations $[A_i]$ is bimodal (Fig.~\ref{fig:bimod}A) and define all $X_{kmi}$ exceeding a threshold, e.g., $1/2$, as active. Another is to first run a deterministic simulation (with $\omega=0$) and only after a given time ($T=10^2$) apply the threshold to define the state. We verify that the two approaches lead to equivalent stable states in most cases (Fig.~\ref{fig:bimod}B) and we adopt the second approach. The similarity between states is defined as the fraction of active or inactive $A_i$ that they have in common. In Figs.~3-5, the state at the end of a trajectory is obtained by taking the median state over the $M$ cells.

The productivity of a state is simply the fraction of its active $A_i$.

To quantify the number of states reached at some time, we calculate the frequency (or prevalence) $f_k$ of each state $k$ and compute the Shannon diversity, defined as $\exp(-\sum_kf_k\ln f_k)$.

\subsection{Figures}

Unless otherwise stated, we fix the following parameters: $N=50$, $\L_i=1$, $\G=10$, $[B]_\tot=0.25$, $\omega=0.1$, $\delta=1$. We discretize the space in $M=10$ or $100$ cells, keeping a constant diffusion $D=1$. Since the dynamics are stochastic whenever $\omega>0$, different sample trajectories are obtained starting from the same initial condition, which we take by default to be $[A_i](x,t=0)=0$ for all $i$ and $x$. A particular graph of inhibitory interactions was chosen, which is typical of the ensemble of regular graph of size $N=50$ and connectivity $c=3$. Results with different graphs from the same random ensemble are shown in Fig.~\ref{fig:more_graphs}.\\

In Fig.~2, $\mathcal{N}=20$ sample trajectories are considered with $M=100$ cells and a total time $T=10^5$, sampled logarithmically in $K=10^4$ time points. One of these sample trajectories is used as an illustration in all panels (see Fig.~\ref{fig:more_samples} for the other 19 sample trajectories). The average productivity and the average number of different states over the 20 samples are also shown in panels C and D.\\

In Fig.~3, $\mathcal{N}=10^5$ sample trajectories with $M=10$ and $T=10^2$ are considered. In panel B, the distribution of similarity is computed using a subsample of $10^4$ sample trajectories. The prevalence of a state is the frequency with which it is obtained as a final result in the $\mathcal{N}=10^5$ samples and is therefore lower bounded by $10^{-5}$ (dotted line in panel D).\\

In Fig.~4, $\mathcal{N}=5.10^4$ competitions are performed, concatenating as input the final compositions at the $M=5$ first positions of two of the $\mathcal{N}=10^5$ sample trajectories of Fig.~3. The simulations are run for a time period of $T=10^2$. For each competition, the state associated with the final evolution, denoted $1{+}2$, is compared with the two states associated with the initial condition, denoted $1$ and $2$. In panel B, the similarity to the closest of these two states is shown in red. In panel C, pairs of initial states are ranked by the absolute value of their difference of productivity, and the fraction of times the final state is closer to the state with the highest productivity is reported.\\

In Fig.~5, $\mathcal{N}=10^3$ sample trajectories are drawn, each with different values of $\L_i$, see Eq.~(1), taken uniformly at random in $[0,1]$. Competitions are then run as in Fig.~4, using the environments in which one of the two initial states was obtained. The distributions of similarities between the final state and the two initial states are shown, in red for the state evolved in the same environment and in green for the other.

\renewcommand{\thefigure}{S\arabic{figure}}

\begin{figure}[b]
\centering
\includegraphics[width=.75\linewidth]{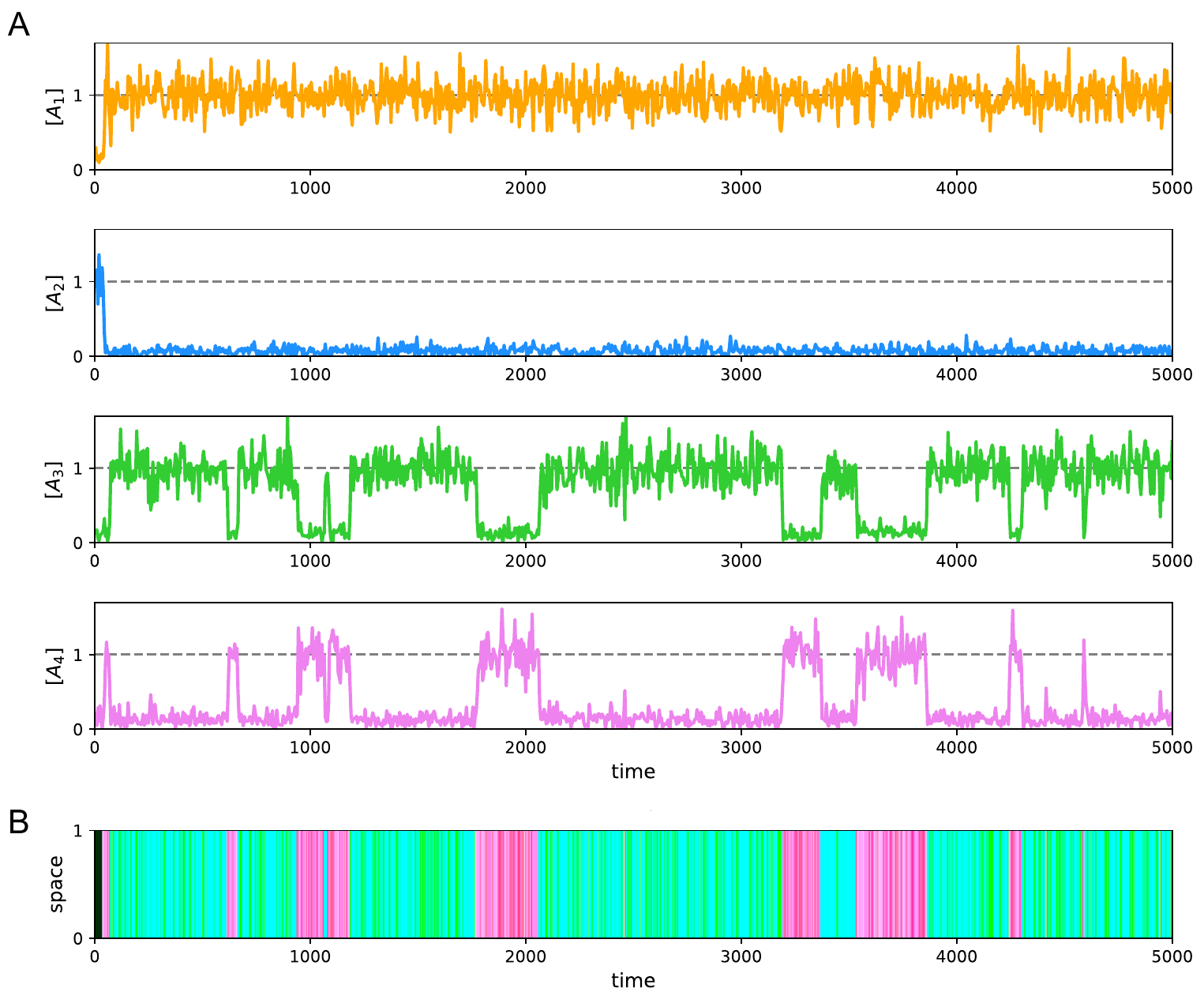}
\caption{{\bf A.} Example trajectory for a well-mixed system with $N=4$ nodes and the inhibition graph shown in Fig.~1C. The initial condition is the state with productivity $P=1/4=0.25$, where the blue node is active and all others are inactive (see Fig.~1D), but the system quickly leaves this state to alternate between the two other states with productivity $P=2/4=0.5$, where the concentrations $[A_i]$ associated with the green and pink nodes alternate between values close to 0 and 1. The parameters are $\G=10$, $[B_i]_\tot=0.25$ and $\omega=0.2$. {\bf B.} Representation of the same trajectory using UMAP, showing the alternation between two states except for the very beginning (the colors of the nodes for $A_3$ and $A_4$ in Fig.~1C are chosen to correspond to the colors of the states where they are at high concentration). The system being well-mixed, the composition is uniform across space.}
\label{fig:toy}
\end{figure}

\begin{figure}[b]
\centering
\includegraphics[width=\linewidth]{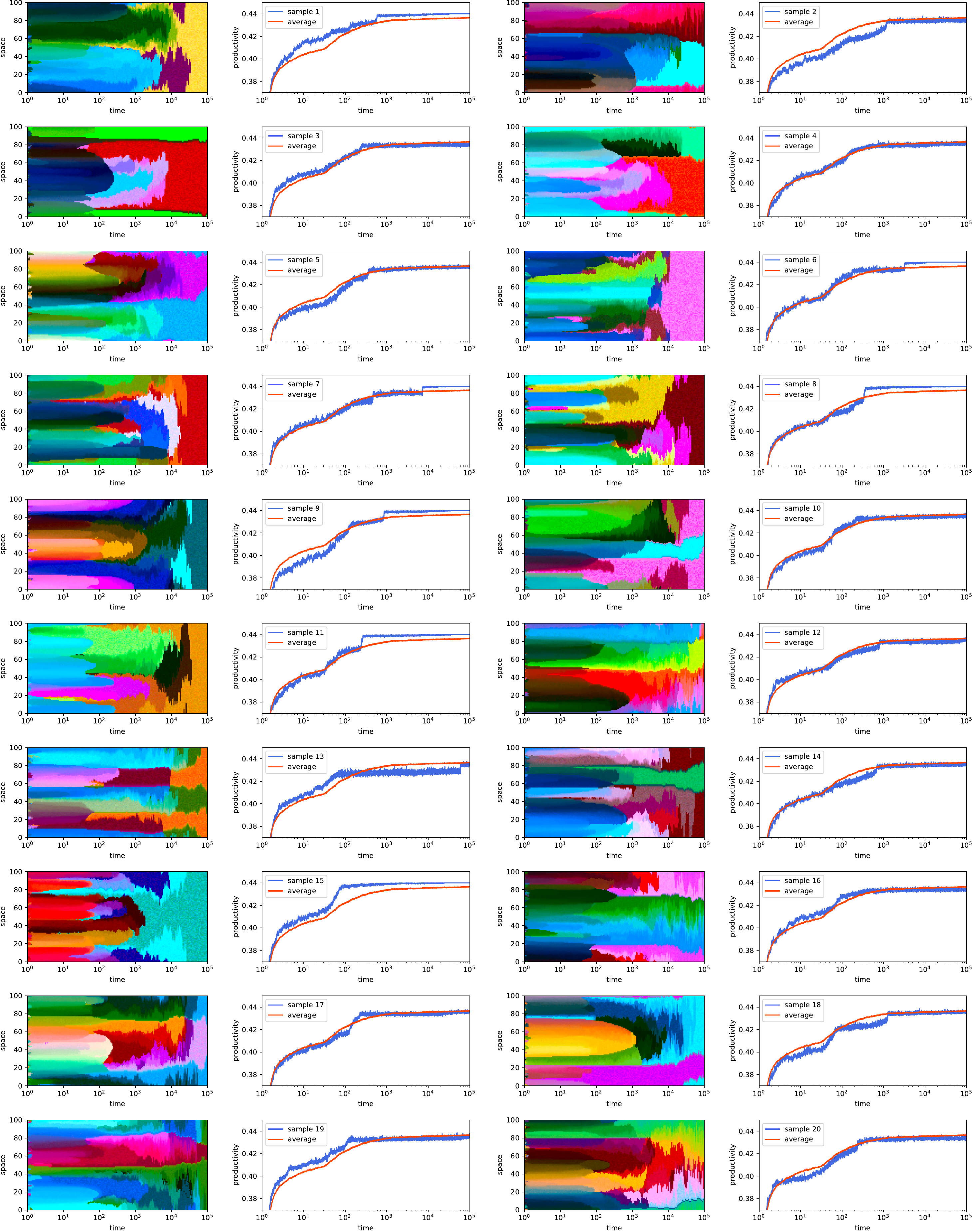}
\caption{20 sample trajectories from the same inhibitory graph as used in Fig.~2 (which shows sample 7). \label{fig:more_samples}}
\end{figure}

\begin{figure}[b]
\centering
\includegraphics[width=\linewidth]{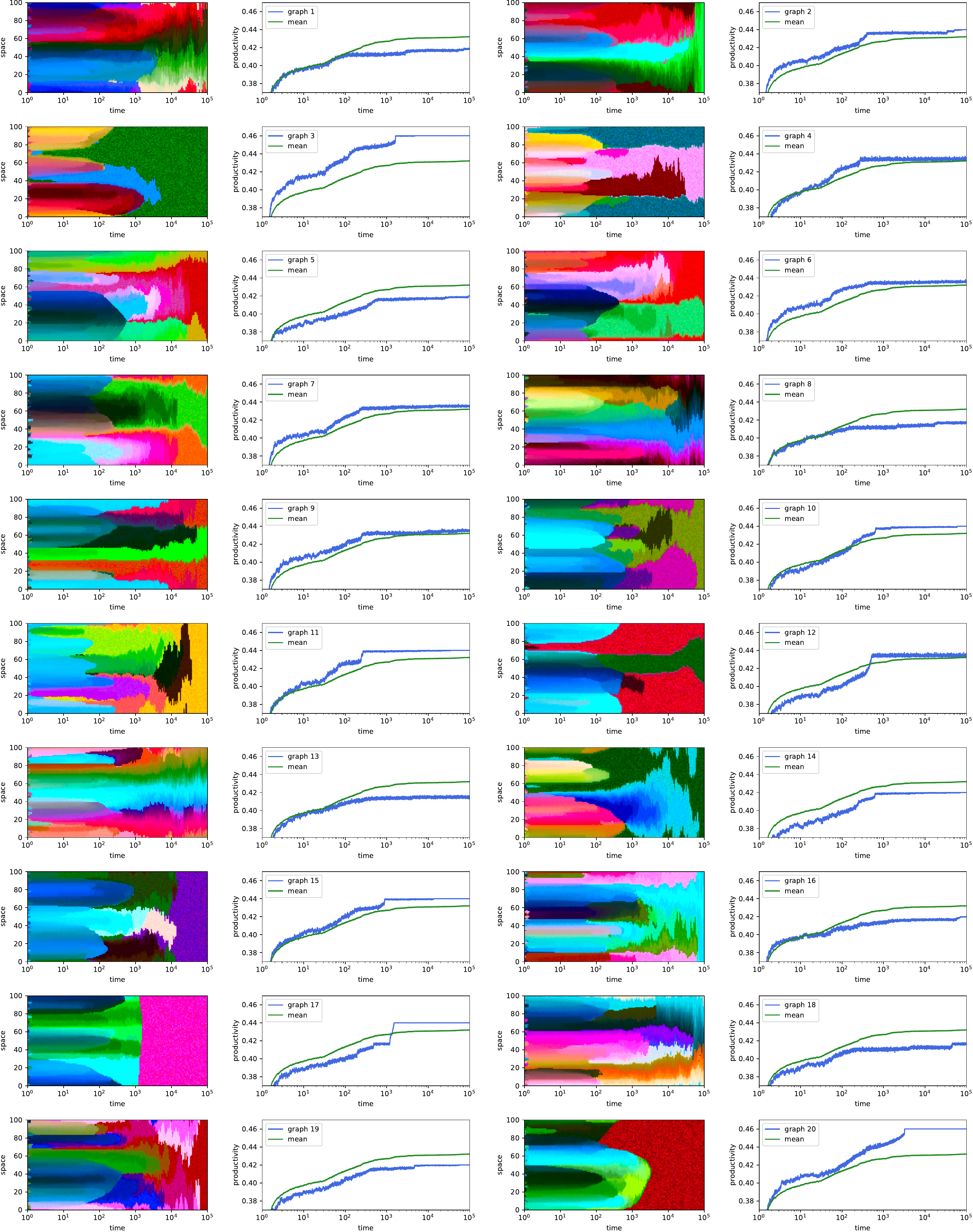}
\caption{20 sample trajectories using different inhibitory graphs (Fig.~2 and Fig.~\ref{fig:more_samples} use graph 11). Note that the scale on the y-axis is different than in Fig.~\ref{fig:more_samples} to account for graphs that allow for higher levels of productivity, e.g. graphs 3 and 20. \label{fig:more_graphs}}
\end{figure}

\begin{figure}[b]
\centering
\includegraphics[width=0.75\linewidth]{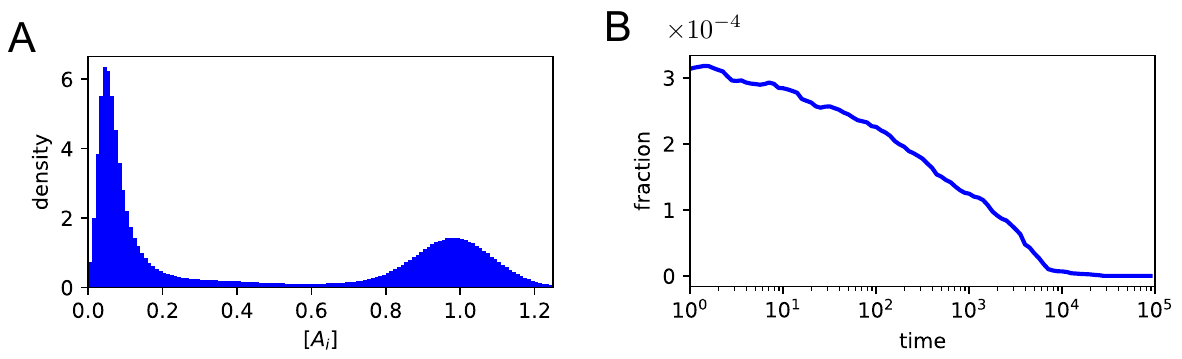}
\caption{{\bf A.} Distribution of concentration $[A_i]$ over space and (log) time for the sample trajectory shown in Fig.~2. The distribution is bimodal, reflecting either active or suppressed $A_i$. {\bf B.} An alternative to classifying as active an $A_i$ with concentration $[A_i]>0.5$ is to use the concentrations of $[A_i]$ as the starting point of a deterministic simulation (without demographic noise) and to identify the stable state to which it necessarily converges. In most cases, the two procedures coincide. The graph shows the fraction of cases where there is a difference, which decreases with time and is in any case of the order of only $0.01\%$. \label{fig:bimod}}
\end{figure}

\begin{figure}[b]
\centering
\includegraphics[width=\linewidth]{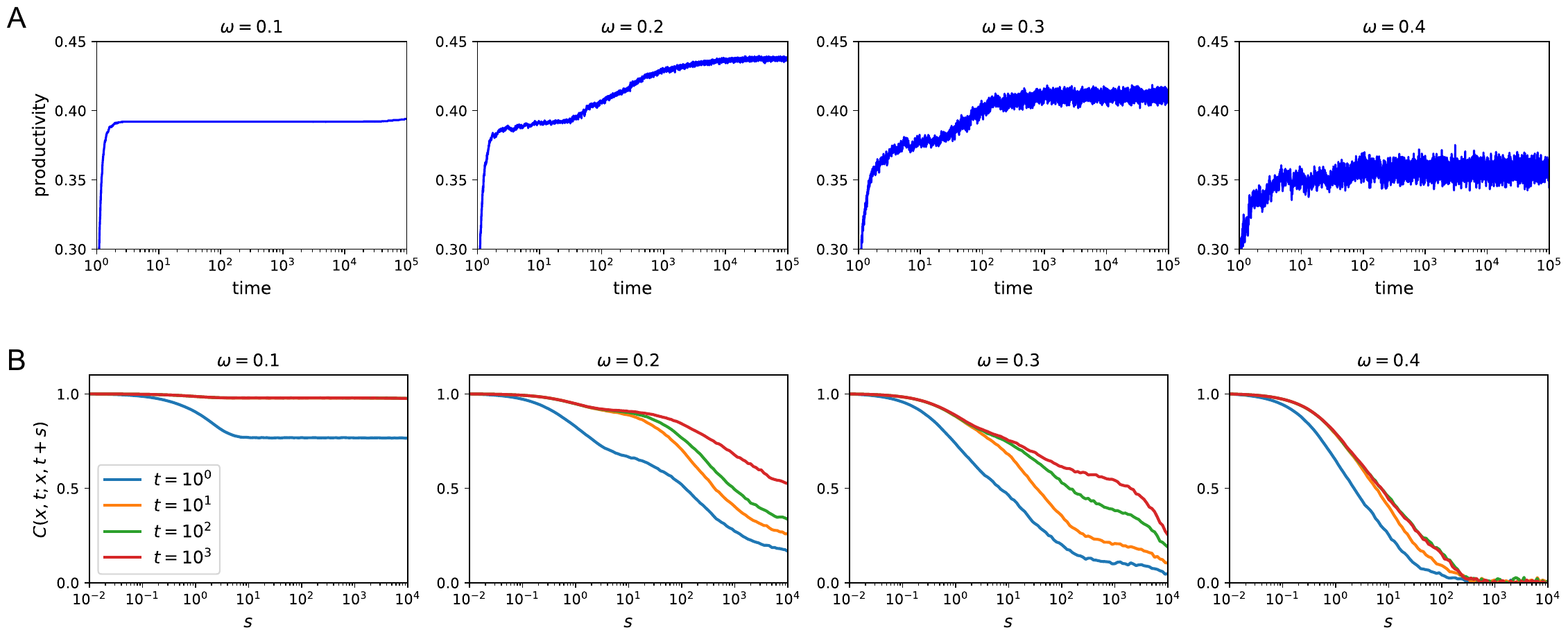}
\caption{{\bf A.} Productivity in the well-mixed case for different values of $\omega$, averaged over 100 sample trajectories. {\bf B.} Temporal correlation functions, showing ``aging'' for small enough values of $\omega$: the correlations grow with increasing values of $t$.\label{fig:D0}}
\end{figure}

\begin{figure}[b]
\centering
\includegraphics[width=0.7\linewidth]{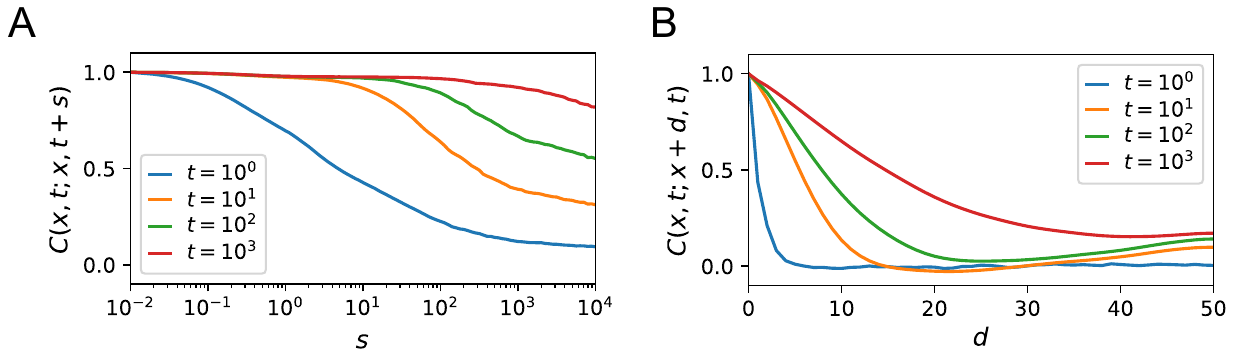}
\caption{Temporal and spatial correlation functions for the 20 trajectories shown in Fig.~\ref{fig:more_samples}. {\bf A.} Temporal correlation function $C(x,t;x,t+s)$ as a function of time interval $s$ for different times $t$, averaged over space $x$ and over the 20 samples. {\bf B.} Spatial correlation function $C(x,t;x+d,t)$ as a function of the spatial interval $d$ for different times $t$, averaged over the 20 samples. The correlations are computed as Pearson correlations using the concentrations $[A_i]$. Both the temporal and spatial correlations increase with time $t$. \label{fig:corrfct}}
\end{figure}

\begin{figure}[t]
\centering
\includegraphics[width=0.25\linewidth]{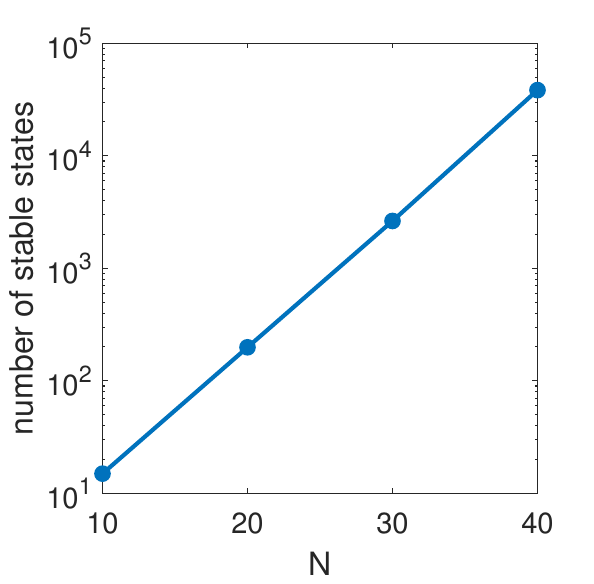}
\caption{The number of stable states grows exponentially with graph size $N$. Here shown for random regular graphs with connectivity $c=3$. Results obtained by averaging over exhaustive counts, using the algorithm of \cite{eppstein2010listing}. Errorbars are smaller than marker size.\label{complex_vs_N}}
\end{figure}

\begin{figure}[t]
\centering
\includegraphics[width=0.25\linewidth]{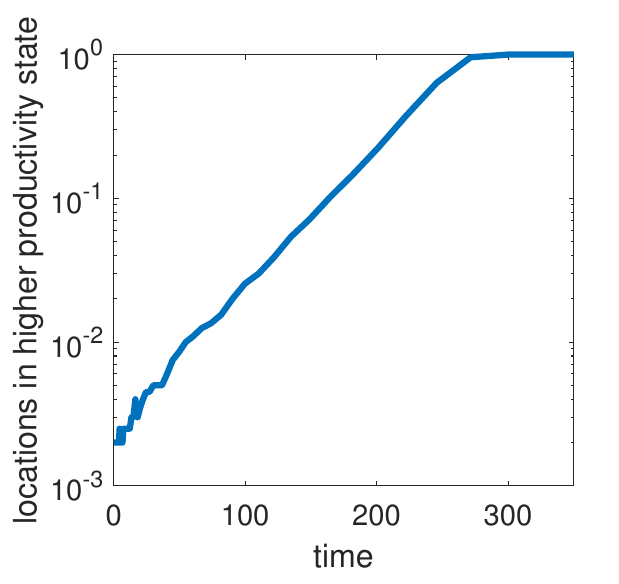}
\caption{Exponential growth in the number of spatial locations in a given state is observed when the connections between spatial locations are ordered in appropriate ways. For example, here migration between locations in space forms a random regular graph with degree 3. The number of variables is $N=4$, and the interacting pairs of variables in a given location are $(1,2),(1,3),(1,4)$. This gives two possible states: only variable 1 is ``on'' and the rest ``off'', or $2,3,4$ are ``on''. Four connected locations (a central location and its three neighbors) in the latter state, and the rest in the former. The number of locations is $M=2000$, and $\omega=0.14$, $B=0.25$, $\Gamma=10$, $D=0.89$.\label{fig:exp}}
\end{figure}

\begin{figure}[b]
\centering
\includegraphics[width=.4\linewidth]{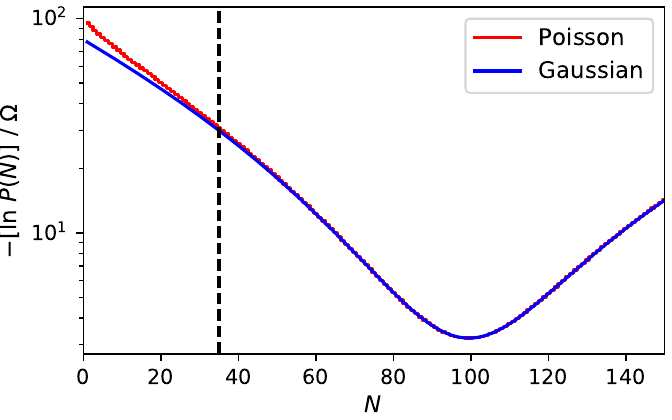}
\caption{
Probability distribution of $N=\Omega [A_1]$ for a single isolated variable $A_1$ subject to Eqs.~\eqref{eq:cg1}-\eqref{eq:cg2} with either Gaussian or Poisson noise. The vertical dashed line indicates the threshold $\Omega (B+1/\Gamma)$ below which the variable would have no inhibitory effect. The parameters $B,\Gamma,\Lambda,\delta$ and $\Omega=1/\omega^2$ are as in the main text. In the range of values of $N$ relevant to the simulations (roughly, to the right of the vertical dashed line), the two distributions coincide up to differences comparable to rounding of the continuous $N$ to an integer. Values where the two distributions differ more significantly (bottom left) are extremely rare ($P\lesssim10^{-20}$).
\label{fig:Gauss_vs_Poiss}}
\end{figure}